\documentclass[12pt]{article}
\usepackage[
top = 2.5cm, 
bottom = 2.5cm, 
left = 2.5cm, 
right = 2.5cm]{geometry}
\pdfoutput=1
\usepackage{color}
\usepackage{graphicx,amssymb,amsfonts,amsmath,amssymb,amscd,amstext, mathrsfs}
\usepackage{xcolor}
\usepackage{graphicx} 
\usepackage{bbm}
\usepackage{dsfont}
\usepackage[colorlinks=true,citecolor=magenta, linkcolor=black]{hyperref}
\usepackage{amsmath}
\usepackage{empheq}
\usepackage{hyperref}

\newlength\dlf

\def\be{\begin{eqnarray}}
\def\ee{\end{eqnarray}}
\def \bea {\begin{equation}}
\def \eea {\end{equation}}

\def\tr{\mathop{\mathrm{tr}}}

\def \nn {\nonumber}

\def \d {\partial}

\def \la {\langle}

\def \rr {\raise.35ex\hbox{\small $\prime$}\kern-.17em{\mbox{\large $\imath$}}}
\def \del {\partial}
\def \dels {\partial\kern-.5em / \kern.5em}
\def \As {{A\kern-.5em / \kern.5em}}
\def \Ds {D\kern-.7em / \kern.5em}

\def \b {\beta}

\def \G {\Gamma}
\def \eps {\epsilon}
\def \m {\mu}
\def \n {\nu}

\def \lam {\lambda}

\def \s {\sigma}

\def\frac#1#2{{#1\over #2}}

\def \ov {\over}
 \def \la {\label} \def \te {\textstyle} \def \b {\beta}
 \def \foot {\footnote} \def \ha {{\te{1\ov 2}}} 
 \newcommand{\rf}[1]{(\ref{#1})}

\def \no {\nonumber}
\def \iffa {\iffalse} 
\def \Bmn {B_{\m\n}}
\def \ed  {\end{document}}
 \def \s {\sigma}

  \def \b {\beta}
\def\nn{\nonumber}

\def \ci {\cite} \def\ci{\cite}
\def \la {\label}
\def \l {\lambda}
\def\foot{\footnote}
\def \ed  {\end{document}}
 \def \td {\tilde}  \def \OO {{\cal O}}\def \ep  {\epsilon}

\def \ha {{\te {1\over 2}}} \def \six  {{\te {1\over 6}}}  \def \tri {{\te {1\over 3}} }
\def \Tr   {{\rm Tr\,}}
\def \hOO  {\hat{\cal O}}
\def \L  {\Lambda}\def \dd {{\rm d}}
\def \dd {{\rm d}}
\def \tA  {\td A}
\def \cG {{\cal  G}}
\def \cTheta {I}
\begin{document}
\thispagestyle{empty}
\begin{flushright} 

YITP-SB-18-08 \\
Imperial-TP-AT-2018-02
\end{flushright}
\today
\begin{center} 
\vspace{0.5cm} 
{\LARGE{{
{\fontfamily{ptm}\selectfont{Self-dual  6d 2-form fields
coupled to  \\\vspace{0.15 cm} 
non-abelian gauge field:  quantum corrections}}
}}}\\
\vspace{0.7 cm}  {Kuo-Wei Huang$^{a, b}$ , \ Radu Roiban$^{c}$ \ and \ Arkady A. Tseytlin$^{d,}$\footnote{Also at Lebedev Institute, Moscow.  
}} \\ 
\vspace{0.7cm} 
{\it
$^a$ C.~N.~Yang Institute for Theoretical Physics,\\
Stony Brook University, Stony Brook, NY  11794, USA}\\
\vspace{0.2 cm} 
{\it 
$^b$ Perimeter Institute for Theoretical Physics, \\
Waterloo, Ontario N2L 2Y5, Canada}\\
\vspace{0.2 cm} 
{\it $^c$ Department of Physics, The Pennsylvania State University, \\
University Park, PA 16802, USA}\\
\vspace{0.2 cm} 
{\it $^d$ Blackett Laboratory, Imperial College, London SW7 2AZ, U.K.}
\end{center}
\
\
\begin{center} 
{\fontfamily{ptm}\selectfont{{Abstract}}}
\end{center} 

We study a 6d model  of  a set of self-dual 2-form  $B$-fields 
interacting with a non-abelian vector $A$-field which is restricted to a  5d  subspace. 
One motivation is that if the gauge vector could be expressed in terms of the $B$-field
or integrated out,  this model could  lead to an 
interacting theory  of $B$-fields only. 
Treating the 5d gauge vector as a background field, we compute the divergent part  of the 
corresponding one-loop effective action which has the $(DF)^2+F^3$ structure  and 
 compare it with  similar contributions from other 6d fields.  
We also   discuss a  4d analog of the non-abelian self-dual model, which  turns out to be   UV finite.

\newpage
\tableofcontents

\setcounter{equation}{0} 
\setcounter{footnote}{0}
\setcounter{section}{0}
\renewcommand{\theequation}{1.\arabic{equation}}

\section{Introduction}

The possible existence  of interacting theories  of non-abelian  2-form fields  in 6 dimensions
possessing some unusual   properties such as  lack of manifest Lorentz symmetry and/or locality is an important 
open problem  (for a recent review  and references  see, e.g., \ci{Saemann:2017zpd}). 
As a first step, one may  study a  system of    2-form potentials  $B$  in  some  representation of gauge group $G$  coupled  ``minimally"  to a non-abelian  
gauge vector $A$.  Similar  couplings appeared, e.g., in the context 
of attempts  to construct  an interacting  theory of  6d (2,0)  tensor multiplets  in  \cite{Samtleben:2011fj} 
  (see also \cite{Lambert:2010wm,Chu:2012um,Bonetti:2012st}).

Here we shall  consider  a  simple  bosonic model of interacting $(B,A)$ fields  following 
\ci{Ho:2011ni,Ho:2012nt, Huang:2012tu}.
We shall study   both the model with self-dual $B$-field strength 
and the  non-chiral $B$-field model.
It turns out that a consistent  gauge-invariant coupling is  possible  provided
  one  keeps    only the 5d  part  of  the 6d   Lorentz  symmetry.\foot{This  may not  be
     unnatural  given  that already   at the free  
level the Lagrangian description of a  self-dual  $B$-field is not manifestly Lorentz invariant.} 
The action is quadratic in $B$ and  takes  a local form in a particular gauge,  with the  $A$-field  restricted to 
 ``live" only  in 5d subspace of the 6d space.
  More generally, 
 one may   attempt   to  consider an extension  where   $A$ is     expressed 
 in terms of $B$  leading to a non-local interacting theory of $B$-fields only.

Our aim  will be to  study   this   $(B,A)$   model    at the quantum level.\foot{The model of \ci{Ho:2011ni}   in the generalized version adopted below has an advantage of 
having an explicit Lagrangian formulation for massless 6d 2-forms without introducing extra auxiliary fields. 
It would be interesting also  to perform a  quantum study of similar models considered in refs. 
\cite{Saemann:2017zpd, Samtleben:2011fj,Lambert:2010wm,Chu:2012um,Bonetti:2012st}.}  
We shall   concentrate  on the one-loop   approximation where $B$ is integrated out and 
 $A$  is treated  as a  background. 
As is well known, quantizing free scalar, spinor or Yang-Mills (YM) fields  coupled to an external vector 
 in  6 dimensions 
 produces $(DF)^2+ F^3$  logarithmic UV divergences in the effective action  (see, e.g., \ci{Fradkin:1982kf}).  
 We shall find that    similar divergences   appear  also  from  the $B$-field loop, 
 implying, in particular,  the breaking of the classical scale invariance. 
 One  may hope to cancel    these  divergences by adding   other fields (e.g., imposing supersymmetry) 
  but so far we did not  find such a finite model. 
 
 As in the case of the  6d Weyl  fermions \ci{Frampton:1983ah,AlvarezGaume:1983ig},
 one could  expect that   the chiral nature  of the self-dual $B$-field  model 
    implies  the presence of 
 anomalous (gauge-symmetry breaking)   terms in the parity-odd  part of the  effective action
 (which   would be  a  gauge-field  counterpart of the familiar gravitational anomaly
 in the case of a single self-dual tensor  \ci{AlvarezGaume:1983ig,Bastianelli:1989hi}). 
 However,    this  does not happen  in the present case:
 as the $A$-field  is  restricted to  5 dimensions,   
 the effective action  has no parity-odd part, i.e. there   is no gauge anomaly as  in   
 any  5d theory.
 
We start in section 2  with a description  of the  gauge symmetries   and the classical  action of the 
$(B,A)$   model  -- both  its non-chiral   version and  the  chiral version with the self-dual $B$-field strength. 
 In the $B_{i6}=0$ gauge   the corresponding actions take   simple form \rf{10}  and \rf{21}. We shall argue  that  the  one-loop effective action  of non-chiral  model  \rf{26}  should be   twice the effective  action  of the self-dual model.\footnote{If the effective action of the chiral model had a parity-odd component, the effective action of the non-chiral model would be twice the parity-even part of the chiral effective action.}
 
The  general  $(DF)^2+ F^3$   structure \rf{82} of the  UV divergent  part  of the 6d  effective action    in a gauge field 
 background will be  discussed in section 3.  We shall summarize  the results  for the 
 corresponding two  coefficients $\beta_2$  and $\beta_3$   for a  collection of 6d  fields (see \rf{833},\rf{835}). 
 
The values of  $\beta_2$  and $\beta_3$    for the self-dual  and  the non-chiral models will be derived in detail
in section 4  by computing the   divergent parts of the $A^2$ and $A^3$  terms in the effective action.  
We shall use   dimensional   regularization  procedure  (applied only with respect to 5-momenta) 
that  preserves   background  gauge invariance.

Some concluding remarks will be  made in section 5. 
In Appendix A  we review  the structure  of the free  $B$-field partition   function. 
Some standard integrals are summarized in Appendix B. 
The same  values of $\b_2$  \rf{414} and $\b_3$ \rf{419} in the self-dual model 
are independently obtained   in Appendix \ref{appC}  from  the $A^6$ term in the effective action.
In Appendix D  we discuss a non-local effective action of a 4d analog of the  6d self-dual $(B,A)$ model.

\renewcommand{\theequation}{2.\arabic{equation}}
 \setcounter{equation}{0}
\section{Non-abelian $B$-field coupled to gauge vector}
\subsection{Gauge symmetry and field strength}

The abelian  antisymmetric  rank 2 tensor  field  has  a familiar 
 gauge symmetry 
\be
\label{0}
\delta B_{\mu\nu}= \d_\m \eps_\n - \d_\n \eps_\m \ .
\ee 
There is  the 
residual gauge symmetry, $\delta \eps_\m = \d_\m \eta$,  which allows one to remove one component from 
$\eps_\m$  and is thus important for the correct degrees  of freedom count.  
A non-abelian generalization  of \eqref{0} should also  admit   some non-abelian  analog of  this 
 residual 
gauge symmetry. 
The abelian gauge-invariant  3-form field strength is $H_{\mu\nu\lambda}= \d_\m B_{\n\lam}
+ \d_\n B_{\lam\m}+ \d_\l B_{\m\n}$. 
To write down a gauge-invariant action in  a non-abelian case, there  should exist a 
generalized field strength  that  transforms covariantly.

It turns out that it is possible  to construct   such a model if one relaxes the condition of 
 6d Lorentz covariance (and locality).
Our starting point  will  be a model 
involving   a 6d 2-form field $B_{\m\n}$  in some representation 
of   gauge group $G$  and a gauge vector field $A_\m$. 
For the simplicity, we assume that both $B_{\m\n}$ and $A_\m$   are taken in the adjoint representation of  
$G$ and use the following notation: 
$D_\m...= \d_\m...+ [A_\m,... ]  ,~
 F_{\m\n}= \del_\m A_\n - \del_\n A_\m + [A_\m, A_\n]$.  
One  can define  the  non-abelian gauge transformations as  \cite{Ho:2011ni}  
\be
\label{1}
\delta A_{\m}&=& D_\m  \lambda \ , \\ 
\label{2}
\delta B_{\m\n}&=& D_\m \eps_\n -  D_\n  \eps_\m- [F_{\m\n},  (n^\rho \partial_\rho)^{-1} (n^\sigma \eps_\sigma)  ]+ [B_{\m\n}, \lambda] \ .
\ee 
Here $\l$ is the parameter of $A_\m$ gauge transformations  under which $B_{\m\n}$  transforms covariantly; 
  $\eps_\m$ is  the parameter of the gauge transformations of $B_{\m\n}$,  which, like $\l$, is now 
  taking 
values in the algebra of $G$. 
The vector $n_\m$  is a fixed constant unit vector which selects a particular direction in 6d space 
breaking $O(6)$ symmetry to $O(5)$. 
In the abelian limit, the gauge transformation \eqref{2} reduces to \eqref{0}.

The structure of the non-local term   in \rf{2}   is  chosen to be  such that 
 if we further assume that  $n^\m A_\m=0$ then there is a non-abelian  generalization 
of the  residual gauge symmetry of the parameter $\eps_\m$   in \rf{0}  
 under which $\delta \Bmn$ is invariant:
\be\la{3}
 \delta \eps_\m= D_\m \eta \ ,\qquad \qquad \delta \lambda=0\ .
\ee 
If we  impose the  additional  condition   that $n^\m \del_\m A_\nu=0$, i.e.   that $A_\m$   depends only on 
5 of the 6 coordinates  (so that, in particular,  $[(n^\m \d_\m)^{-1}, D_\n] f=0$  for  a 6d function $f(x_\lambda)$)  
  then one can check that  the gauge algebra  closes:
\be\la{4}
[\delta_1, \delta_2] = \delta_3\ , ~~~\rm{with}~~~\quad
\lambda_3= [\lambda_1,\lambda_2]\ ,~~\quad  \eps_{\m 3}=[\lam_1,\eps_{\m 2}]-[\lam_2,\eps_{\m 1}]\ .
\ee
The corresponding  field strength of $\Bmn$ is  defined  as  
\be
\label{H}
&&H_{\m\n\lam}= D_\m B_{\n\lam}+D_\n B_{\lam\mu}+D_\lam B_{\m\n}\nn\\
&&\ \ \ \ \ \ \quad \ \ + [F_{\m\n}, (n^\rho \partial_\rho)^{-1} (n^\sigma B_{\lambda\sigma}) ]
+[F_{\n\lam}, (n^\rho \partial_\rho)^{-1} (n^\sigma B_{\m\sigma})]+[F_{\lam\m}, (n^\rho \partial_\rho)^{-1} (n^\sigma B_{\n\sigma})]\ ,  \ \ \ \ \ \ 
\ee 
where the non-local terms   ensure  that  $H$  transforms covariantly:  
\be
\delta H_{\m\n\s}= [H_{\m\n\s}, \lambda] \ . \la{HH}
\ee 
Thus 
one can consistently couple the non-abelian antisymmetric tensor  to a non-abelian gauge 
field restricted to a codimension-1 (``boundary")  subspace, i.e. with an 
effective  non-locality  along the ``bulk" direction (see Fig.1). 
This non-locality  may be viewed as  a gauge artifact  as there is a gauge in which the corresponding 
action is local (see below). 
Note also  that we  do not impose any  boundary condition at $x_6=0$.

\begin{figure}

\begin{center}
\includegraphics[width=2 in]{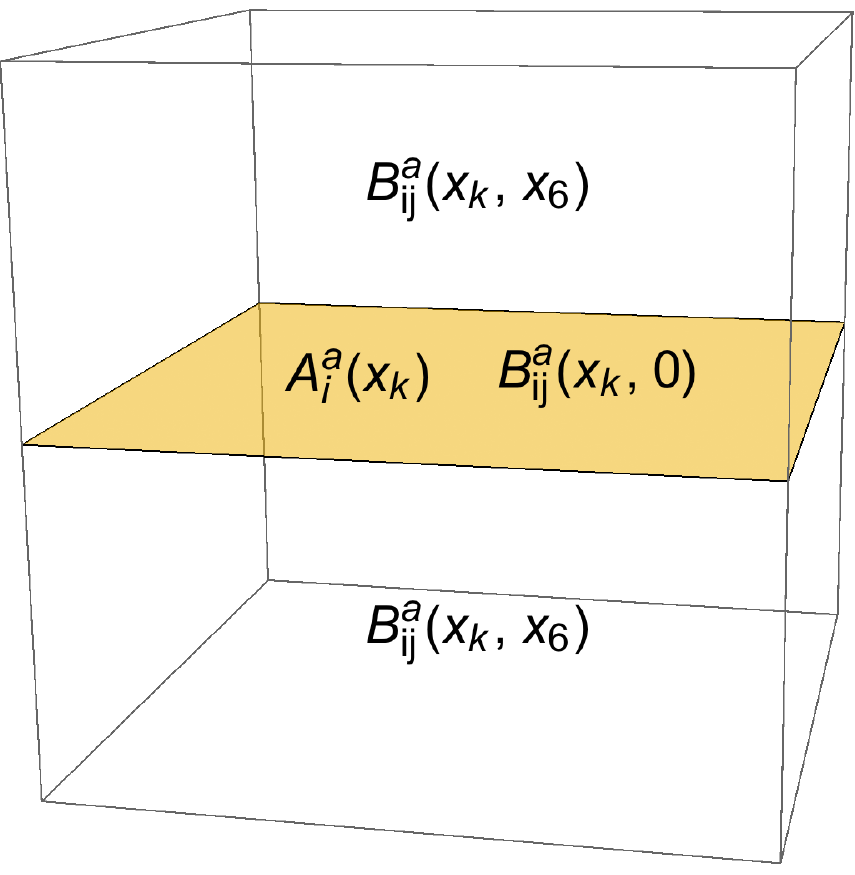}
\end{center}
\caption{\small{
Sketch of $B$- and $A$-fields in 6d space.
The $B$-field has 5-indices (in $B_{i6}=0$ gauge)  but depends on all 6 coordinates. The $A$-field ``lives"  only  in a codimension-1 subspace with $x^6=0$ (colored region) where the interaction  takes place. 
}
}
\end{figure}

Without  loss of  generality, one  
 can always  choose   $n_\m$ to point in the 6th   direction, i.e.  $n_\m=(0,0,0,0,0,1)$,   so 
 that   the vector field restricted as above ``lives'' in 5d subspace\foot{We 
  assume the Euclidean signature with 6d indices $\mu, \nu, \lambda,...=1, ...,6$ and use 
$ i, j, k,..=1, ...,5$  for 5d indices.} 
 \be \la{5} 
 A_\m= \{A_i(x^k, 0),\ A_6=0\} \ , \quad   F_{i6}=0\ , \quad\ \ 
 F_{ij} = \del_i A_j - \del_j A_i + [A_i, A_j] \ , \quad   D_6 = \del_6 \ .  
\ee
This  $(B,A)$ model may be viewed as an intermediate step towards constructing an interacting model of $B$-fields only. 
For example, one  may  interpret  $A$   not  as an independent field but   as  related to $B$  by some non-local condition. 
In \cite{Ho:2011ni}  the $x^6$    direction was  assumed  to be compactified  to a circle of radius $R$ 
 and $A_i$  was related to the zero mode of $B_{i6}$: 
$A_i\equiv   \int  dx^6 \,  B_{i 6} = 2 \pi R   B^{\rm{(zero~mode)}}_{i6}(x^k)$. 
Moreover, the zero mode of the 3-form field strength was  defined directly via the Hodge duality: 
$H_{ijk}(x^k,0)\equiv  
 {i \over 4 \pi R}  \eps_{ijkmn} F_{mn}$.\foot{Note that  this implies $\int d^6x \, H^2_{ijk}(x^k,0) \to {1\over R} \int d^5 x \, F^2_{ij}$, which is  formally consistent with
scaling symmetry.   
The compactification assumption  naturally breaks  the global  $SO(6)$ 
symmetry to $SO(5) \times SO(2)$.
In the non-compact  $x^6$ case  one may set 
$
A_i \equiv 
 \int dx^6 \,  B_{i 6} , \ \ 
\lambda \equiv  
\int dx^6 \,   \eps_6 
$  and impose  the boundary  conditions: $\eta (x_i, \pm \infty)=0$, \ $\eps_i (x_k, \pm \infty) = [A_i, \lambda]$, to preserve the gauge-covariant structure. 
}
One may  also 
 treat $A_i $  first as an independent quantum field and then 
  integrate  it  out  in the path integral  obtaining 
an effective  non-local  model of self-coupled  non-abelian  $B$-fields.

Below   we shall  view   $A_i$  just   as a background   field 
coupled  to the quantum $B$-fields.  
This interacting  theory  will   have only  $SO(5)$ part  of the  full  6d rotational (Lorentz) symmetry.

\subsection{Classical action and gauge fixing}

Our starting point  will be the following gauge invariant action  describing the non-abelian 
 6d field $\Bmn (x^\m) $  coupled  to the 5d gauge field $A_i (x^j)$:
\be
\label{6}
S_{ }= \six  \int d^6x ~ \Tr (H_{\m\n\lambda}H^{\m\n\lambda}) \ .
\ee  
Here $\Tr$ is  in some representation $R$ of  gauge group,   with 
$\Tr (H^a t^a)^2 = T_R  H^a  H^a$ ($a=1, ..., \dim G$),   with  $t^a$   being hermitian generators  and 
$T_R= \ha $  or $T_R= C_2$  
 if $H$ is a matrix  with indices  in fundamental or adjoint representation. 

The  overall  (dimensionless) normalization  constant in  the action \rf{6} 
  will   not  be   important as in this paper we will only consider 
the 1-loop approximation treating  $A_i$ as a background field. 
In general, to make the model renormalizable one  would need to introduce also $A$-dependent counterterms 
 $\int d^6x \, [c_1 (DF)^2 + c_2  F^3]$ (see below), i.e. two  extra dimensionless  
   coupling constants.\foot{In general, one  could  also   consider  adding the  5d Chern-Simons  action for $A$:

$
S_{\rm{5d}}={\kappa\over 3}\int \Tr (A \wedge F \wedge F
+ {i \over 2} A \wedge A \wedge A \wedge F
- {1\over 10}A \wedge A \wedge A \wedge A \wedge A),
$
but it  will  not be naturally induced in the  model  based on \rf{6} and also in its self-dual version
discussed below.}

From \rf{H}   and  \rf{5} we have 
\be \la{7}
H_{ij6} = \del_6 B_{ij}  +D_i B_{j6} - D_j B_{i6}  \ ,\qquad  ~~ H_{ijk} =D_i B_{jk} 
 +  [F_{ij}, \del_6^{-1}  B_{k6} ]   + { (i,j,k \rm \ cycle)} \ . 
\ee
The action \rf{6} is invariant under gauge transformations \rf{1} and \rf{2}, i.e. 
\be
\label{8} 
\delta B_{ij}= D_i  \bar \eps_j -  D_j \bar \eps_i +[B_{ij}, \lambda] \ , ~~~~ \delta B_{i6}= -\partial_6 \bar \eps_i+[B_{i6}, \lambda] \ , ~~~~\ \ \ \   \delta A_{i}=  D_i \lambda \ , 
\ee  
where   $\l=\l(x_i)$   and   we redefined the gauge parameter $\eps_i\to \bar \eps_i $ as 
\be\la{88}
 \bar \eps_i=  \eps_i- D_i \partial_6^{-1} \eps_6 \ .
\ee
We can fix the   $\bar \eps_i$ gauge freedom  by the natural gauge $B_{i6}=0$  
in which the field   strength becomes manifestly local 
\be 
\la{9}
B_{i6}=0 : ~~~~\ \ 
H_{ijk}= D_i  B_{jk} + D_j B_{ki} + D_k  B_{ij} \ , ~~~~\ \ \ 
H_{ij6}= \d_6 B_{ij}  \ .
\ee  
The gauge-fixed action  \rf{6} is  then  given by
\be
\label{10}
&& S_{ }= {\te {1\over 2}} \int d^6x ~ \Tr \big[
  ( \del_6 B_{ij} )^2    +      \tri  H^{ijk}H_{ijk }  \big]  
 =  \ha \int d^6x ~ \Tr  \big( B^{ij}  \Delta_{ij}^{mn} B_{mn}\big) \ , 
\\
&& \Delta_{ij}^{mn} = -\delta^{mn}_{ij}(\del_6^2 + D^2)  +  2\delta^{[m}_{[{i}} D^{n]} D_{j]}
\ , \qquad\quad  D^2\equiv D^iD_i\ , \quad 
 \delta^{mn}_{ij}= \delta_{[i}^m \delta_{j]}^n  \ ,
\la{11}\ee
where $[ij]$   stands for antisymmetrisation with weight $\ha$   and  
$D_i  B_{jk} \equiv    \del_i B_{jk} + [A_i, B_{jk}]$.\foot{Note  that the   6d action \rf{6} or \rf{10}   is  manifestly  scale invariant. 
Starting   with such   an   $(d B + A B )^2  $   action 
 and    integrating out  $A_i $    should  give a local, scale-invariant 
  but non-polynomial   and non-Lorentz-invariant 
   action  $\sim (BB)^{-1} dB dB$  for  the   non-abelian $B$-fields. 
 It  will not have free  quadratic  part and will  thus require some  non-trivial 
 (scale-invariance breaking)  $B$-field background 
  to define  a perturbation theory (cf. \cite{Beccaria:2017aqc}).
  }

Our aim will  be  to compute the  logarithmic divergences in  the $A_i$-dependent 
1-loop  effective action  found by integrating out the $B$-field in \rf{10}\foot{We  ignore 
the $A_i$-independent  factors in the partition function $Z=e^{-\G}$  that  should   agree with 
\rf{a5} in the  free  limit. See Appendix \ref{appA} for a discussion of the free  $B$-field partition function  
  in the ``axial" gauge $B_{i6}=0$.
}
\be 
\la{12} \Gamma= \ha  \log \det \Delta_{ij}^{mn} (A)  \ . 
\ee 
The  operator $\Delta_{ij}^{mn} $ in \rf{11} defined on 6d field $B_{ij} (x^\m) $ is, in general,  non-degenerate  
 and   the gauge  condition  $B_{i6}=0$   does not  lead to a non-trivial  ($A$-dependent)
  ghost  determinant (cf. \rf{8}).
Note that the gauge-fixed action \rf{10} is still invariant under the following 5d local gauge transformations  ($ U(x^i) \in G$):
\be
\label{13}
B'_{ij} = U B_{ij} U^{-1} \ , ~~ \qquad A'_i = U A_i U^{-1} +  U \partial_i U^{-1}  \ .
\ee 
Provided  the regularization preserves   this symmetry, the effective  action \rf{12} 
should  thus be  built out of gauge-invariant   combinations of 
$F_{ij}$  and $D_i$.  

\subsection{Self-dual $B$-field model  }
 
Let us  now consider the analog of the non-abelian action \rf{10}  in the case of  the  
 $B$-field  with  a self-dual field strength. 
Let us  first  review the   free-field   case of a single   self-dual   field. 
In Minkowski    signature the real 6d self-duality condition reads
\be\la{14}
H_{\m\n\lambda}= 
 {\six} \eps_{\m\n\lambda\sigma\rho\delta}H^{\sigma\rho\delta}  \ .
\ee 
As is  well known,  one   way to   find the action   corresponding to  \rf{14}  is to relax  the 
manifest Lorentz   symmetry.  A systematic approach is to start   with the phase-space path integral   for the non-chiral $H^2_{\m\n\l}$   theory, impose the   standard  
 ``time-like"   gauge  $B_{i0}=0$,  trade the momenta  corresponding to $B_{ij}$   for another  2-form  field    and then impose 
the self-duality truncation   ending up with  the ``$  {\cal E B - BB}$"  type action ($\cal E$ is  ``electric"   and $\cal B$ is ``magnetic")  \cite{Henneaux:1988gg} (see  also  \cite{Schwarz:1993vs,Perry:1996mk,Pasti:1996vs,Belov:2006jd}).\foot{To  get the right  count of degrees of freedom  at the level  of path integral   one should  also   keep track of   appropriate  Jacobians in the path integral  measure.}
Switching to the Euclidean notation 
that we shall use   below ($x^0 \to i x^6$,  with the gauge condition $B_{i6}=0$)
   the resulting action is  ($i,j, ...=1, ..., 5$)
\be \la{15} 
\td S_{+} = \int d^6 x ~   \, \ha i \, \ep_{ijkpq} \del_k B_{pq}
 \big( \del_6 B_{ij}   + \ha i \,  \ep_{ijrmn} \del_r B_{mn} \big) 
\ .  \ee
It  formally  has a  residual   gauge invariance
$\delta  B_{ij}  = \del_i  \xi_j - \del_j \xi_i $.\footnote{This a 6d 
 symmetry; the action is invariant up to a surface term. 
This residual symmetry is an artifact of the action \rf{15} -- it is   absent   in  the
required  self-duality equation $ \OO_+  B_{ij} =0$.}
Taking the  variation of \rf{15}  over $B_{ij}$  we obtain
 the equation of motion 
which may be written as\footnote{As in \rf{11}, we  adopt the standard convention $
\delta^{mn}_{ij}= {1\over 2}(\delta_i^m \delta_j^n-\delta_i^n \delta_j^m)$.}
\be \la{16}
  \del_{[k}  \OO_+  B_{ij]} =0\ , \qquad \qquad 
(\OO_\pm)_{ij,mn}   \equiv   \delta_{ij, mn}   \del_6   \pm  \ha i \,  \ep_{ij rmn} \del_r \ . \ee
  It is solved by 
\be 
\la{166} \OO_+  B_{ij} = \del_i  q_j (x_i) - \del_j q_i(x_i) +  f_{ij} (x^6) \ , 
\ee 
for a 5d function $q_i (x_i)$ and a function $f_{ij} (x^6)$ that does not depend on 5d coordinates. 
Absorbing $q_i$ part into a formal redefinition of $B_{ij}$  in $\del_6 B_{ij}$ term in 
$ \OO_+  B_{ij} $   and 
  imposing the boundary condition that   the self-duality condition 
$\OO_+  B_{ij}=0$ is satisfied  at  ``spatial infinity" $|x^i| = \infty$  
 we conclude  that   $f_{ij}=0$   and thus $\OO_+  B_{ij}=0$
 is   satisfied  everywhere.   
 Integrating  over $B_{ij}$  in the path integral 
defined by the   action  \rf{15}  and taking into account  the  necessary determinant 
 factors in the measure  one finds  that the resulting partition function  is 
 \be \la{17}
 Z_+ = \big(\det \OO^\perp_{+}\big)^{-1/2} \ , \ee  
 where   $\OO_{+}^\perp$  acts on  transverse $B^\perp_{ij}$   field 
  and thus describes 3 dynamical degrees of freedom 
 as expected.\foot{$B^\perp_{ij}$   has   $ \ha \times 4 \times 5 -  (5-1) =6$ 
  real  components  and that  the differential operator is a 1-st order one
  (cf.  Appendix \ref{appA}).}

The equivalent   results   can be   obtained   by starting with  an alternative  (``$\cal  E B -  EE$")
  action  
\be 
\la{18} 
S_{+} = \int d^6 x   \, \del_6   B_{ij}   \big( \del_6 B_{ij}   + \ha i \,  \ep_{ijkmn} \del_k B_{mn} \big) 
\ .  \ee
Here  the  equations of motion  $\del_6 (\OO_+  B_{ij})=0$     reduce to 
$ \OO_+  B_{ij} =  f_{ij} (x^k), $    and thus  if  the self-duality condition 
$\OO_+ B_{ij} =0$  is imposed  at  $|x^6|=\infty$, it is satisfied everywhere. 
The action \eqref{18} has a 5d residual gauge symmetry $\delta  B_{ij}  = \del_i  \xi_j - \del_j \xi_i$, where $\xi_i=\xi_i(x^k)$.  
The $B_{ij}$ path  integral  measure here should  have an extra 
   factor  of $(\det \del_6)^{1/2}$ that ensures 6d Lorentz invariance; 
 as a result one finds the same chiral partition function \rf{17} (cf. Appendix  \ref{appA}). 

Note that  the free limit of the  non-chiral action \rf{10}  is equivalent 
to a combination of the self-dual and anti self-dual models: in the free limit the kinetic term   in \rf{10},\rf{11}  
may be written as 
\be \la{19}
( \del_6 B_{ij} )^2    +      \tri  H^{ijk}H_{ijk } = \OO_+ B_{ij} \, \OO_- B_{ij} \ , \ \ \ \ 
\ee
and thus the   corresponding partition function    is given by 
\be \la{20}
Z= (\det \Delta_\perp)^{-1/2} = Z_+  Z_- \ . \ee

The above  discussion  has a  straightforward 
generalization  to the   non-abelian case.
Namely, let us    require that the self-duality condition 
\rf{14} or its Euclidean  counterpart in the $B_{i6}=0$ gauge 
 should   be  satisfied for the non-abelian field strength 
\rf{H} or \rf{7}, i.e.
\be  \hOO_+  B_{ij}=0  \ , \qquad \qquad \la{23} 
(\hOO_\pm)_{ij,mn}   \equiv   \delta_{ij, mn}   \del_6   \pm  \ha i \,  \ep_{ij kmn} D_k (A) \ . 
\ee
  One may 
expect   that this    condition  should  follow (under  the corresponding 
 boundary conditions  discussed  above) from  the  direct   analogs of \rf{18}  and \rf{15}:
 \be \la{21} 
&&S_{+} =  \int d^6 x \, \Tr\Big[ 
  \, \del_6   B_{ij}   \big( \del_6 B_{ij}   + \ha i \,  \ep_{ijkmn} D_k B_{mn} \big) \Big] \ , 
\  \\
&& \la{22}
\td S_{+} =  \int d^6 x \, \Tr\Big[ 
  \,      \ha i \,  \ep_{ijrpq} D_r B_{pq}      \big( \del_6 B_{ij}   + \ha i \,  \ep_{ijkmn} D_k B_{mn} \big) \Big] \ .
\  
 \ee
 This is indeed  obvious  for \rf{21} 
 but   is    not  immediately  so    for the second action \rf{22}.
 The   equations of motion  following from \rf{22}, $D_{[i} \hOO_+ B_{jk]}=0$,   may be solved  as 
 $\hOO_+ B_{ij} =  q(x^6)  F_{ij}(x^k) +   f_{ij} (x^6)$,    where $F_{ij}$   is the field  strength 
 of $A_i$.   One   may then   attempt to  absorb  the  $F$-term  by a (non-local in $x^6$)
 redefinition   $B_{ij} \to B_{ij} + (\del_6)^{-1} q(x^6)  F_{ij}(x^k)$  to arrive at the self-duality condition. 
 However,  the quantum equivalence of the theories   based on \rf{21} and 
 \rf{22}    becomes unclear   as   an  extra determinant 
 of the operator $  \ha i \,  \ep_{ijrpq} D_r$  coming from \rf{22} 
  will  now have a non-trivial dependence on $A_i$.
  
 In  what follows   we shall   use  the   simplest    
 action \rf{21}  as defining  the  non-abelian self-dual $B$-field  model.\foot{Like the free action \rf{18} the interacting   action \rf{21}  still has 
  the residual 5d gauge symmetry $\delta B_{ij}=\del_i \xi_j- \del_j \xi_i$, $\xi_i=\xi_i(x^k,0)$
  under which  the  variation  of \rf{21}   is  a  total  derivative:
  $\delta S_{+}=  \Tr \int d^6 x \,  
  \, \del_6   B_{ij}   \big( i \,  \ep_{ijkmn} [A_k, \del_{m} \xi_n ]\big)
=\Tr  \int d^6 x \,
  \, \del_6   \big( B_{ij}    i \,  \ep_{ijkmn} [A_k, \del_{m} \xi_n] \big) $.
 Since  the parameter  does  not depend on $x^6$,  this does not imply a degeneracy of the  resulting kinetic operator 
 for generic values of 6-momentum  and thus does not require gauge fixing. 
  }  
 Since the  $\del_6$   operator   factorizes   in \rf{21},  the  corresponding 
 partition function   is   given  by the   direct analog of \rf{17}   with  
 $\OO_+  \to  \hOO_+(A)$.\foot{
 As in \rf{12}  we shall ignore  constant $A$-independent factors in $Z$:  the  
 operator $ \hOO_+$  is  acting on the  full $B_{ij}$ rather  than  on its  transverse part as in the free case in \rf{17}.
 }
  It  is straightforward to check    that the operator $\Delta(A)$ in the non-chiral action \rf{10},\rf{11}  is given again  by the product of the self-dual and anti  self-dual operators in \rf{23}:
 \be \la{24}
 \Delta^{mn}_{ij} (A) =- \hOO^{\ mn}_{+\, pq}(A)\  \hOO^{\ pq}_{-\, ij}(A)  \ . 
 \ee
 As a result, the  non-chiral  $B$-field   quantum effective action  \rf{12} 
 may be written as 
 \be \la{25}
 \Gamma = \Gamma_+  + \Gamma_- \ , \ \ \qquad \qquad 
 \Gamma_\pm  = \ha \log \det \hOO_\pm (A) \ . \ee
 While $\Gamma$ should   be parity (P) even, $\Gamma_+$   and $\Gamma_-$  may a priori  contain imaginary 
 P-odd parts   that cancel in their sum in \rf{23}  (as, e.g., in the   case of  an 
 external gravitational field  \ci{AlvarezGaume:1983ig}). 
 However, it is easy to see that  this does not happen in the present case   when  the external  field $A_i$  does not depend on $x^6$.  Indeed, 
 $\del_6 \to - \del_6$  combined with $\eps_5 \to - \eps_5$ is 
 a symmetry of  the  classical   action    \rf{21}  and thus    should    be present 
 also in the effective action. As the  P-odd  part of $\Gamma_\pm$ 
 should  contain an odd number of 
 $\eps_5 = (\eps_{ijkmn})$ factors  it should  thus have 
 an odd number of   $p_6$  factors  (in momentum representation) 
    but then the integral over  $p_6$ vanishes.
    The absence of P-odd  part    implies also the absence of an 
  anomalous  (5d gauge symmetry breaking)   part of $\G_\pm$. 
  Thus  both  the  effective  action $\G$  of the full  non-chiral theory   and  $\G_+$  of the  self-dual theory 
  should  be invariant under the residual gauge symmetry of the  $A$-field in  \eqref{13}. 

To conclude, we   have
\be 
\la{26}
 \Gamma = 2 \Gamma_+   \ , \ \ \qquad \qquad 
 \Gamma_+ =\Gamma_- =  \ha \log \det \hOO_+ (A) \ . 
\ee

\renewcommand{\theequation}{3.\arabic{equation}}
 \setcounter{equation}{0}
 \section{Structure  of divergent part of  effective   action } 

Before describing the details  of the computation of the  divergent part $\G_\infty$ 
 of the effective action 
\rf{12}  corresponding to non-chiral  non-abelian   $B$-field action \rf{10} 
and the    self-dual  model \rf{21}   and verifying their  relation in  \rf{26}, 
let us   first   discuss the   general   structure  of  $\G_\infty$
in  a   background  gauge vector field.  

Let us  consider the   1-loop  effective action  for a 
  6d  model containing  standard  2-derivative 
 Yang-Mills  vectors, scalars   and  spinors   coupled to a 
 background gauge field. To  prepare for the 
 discussion of the   models in the previous section  we  will  specify  to the 
  case  when  the background  field is chosen to be  the  5-dimensional one 
as  in \rf{5} (i.e.   use indices  $m,n, ...=1, ..., 5$). 
    Using, e.g., the  heat kernel representation  and proper-time cutoff  $\epsilon= \Lambda^{-2}\to 0 $ 
  one  finds  \cite{Fradkin:1982kf}  
  from the general expression  for  the corresponding heat kernel coefficient  
   \cite{Gilkey:1975iq} (see also \cite{Bastianelli:2000hi,Osborn:2015rna}) 
  \be 
 \label{820}
&& \Gamma_\infty = - B_6 \log \Lambda  \ , \  \ \ \ \ \ \\
 &&B_6 = { 1\ov (4 \pi)^3} \int d^6 x \Big[ \te - { 1 \ov 60} \beta_2 \tr (D_m F_{mn} \,  D_k F_{kn})   + { 1 \ov 90} \beta_3 \tr (F_{mn}   F_{nk} F_{km})\Big]  \ . \la{82}
 \ee
 $\beta_2$ and $\beta_3$ are   numerical coefficients of the  two  independent 
  dimension-6    invariants built  out of  the background field.\foot{The other  two  invariants of the same dimension are related by  use of  Bianchi  identities: 
   
   $ \tr (D_m F_{kn} \,  D_m F_{kn}) = 2 \tr (D_m F_{mn} \,  D_k F_{kn})  - 4  \tr (F_{mn}   F_{nk} F_{km}) +{\rm  total \ derivative}\ , $ 
   
   $
   \tr (D_m F_{kn} \,  D_k F_{mn})  = \ha \tr (D_m F_{kn} \,  D_m F_{kn}) . $}
   Note that  in dimensional regularization  one gets $\Gamma_\infty =  {1\ov \dd-6} B_6 $
 where    ${1\ov \dd-6}$  corresponds to  $-\log \L$  in \rf{820}.
Here $\tr$  is over the  matrix  indices of the particular  representation to which the quantum field   belongs:
if it is in the adjoint representation  (with hermitian generators  $(t^a)_{bc}=  -i f^a_{\  bc} $) 
  one has  the gauge field as a  matrix $A^{ab}_n = f^{acb} A^c_n$  and  
\be 
\la{822}
\tr (D_m F_{mn} D_k F_{kn})  &=&  - C_2   D_m F^a_{mn}  D_k F^a_{kn} \ , \qquad  \qquad f_{acd} f_{bcd}=C_2 \delta_{ab}  \\
\la{823}
\tr (F_{mn}   F_{nk} F_{km}) &=&    -  \ha C_2    f^{abc} F^a_{mn}   F^b_{nk} F^c_{km} \ 
.
\ee
For generic  representation $R$ 
with   generators   satisfying  $\tr ( t^a t^b) = T_R   \delta^{ab} $
  one is to replace $C_2$  in \rf{822},\rf{823}  by $T_R$. 
  
For a collection of   $N_1$   6d YM vectors, $N_0$   real scalars   and $N_{1\ov 2} $ Weyl fermions,   each   in adjoint representation,  one finds 
 \cite{Fradkin:1982kf}\foot{We use this opportunity to correct two unfortunate    misprints   in \cite{Fradkin:1982kf}:  
 $d- {1\ov 2} \to d-42 $ in eq. (3.9)   (here $d=6$)
  and  $- {1\ov 72}  \to +  {1\ov 90} $ in eq. (3.5)  (results  in eq. (3.6) there  are correct).}
\be \la{83}
\b_2 = - 36 N_1 +   N_0  + 16 N_{1\ov 2} \ , \ \ \ \ \ \ \qquad\ \ \ 
\b_3 = 4 N_1 +   N_0  -  4 N_{1\ov 2}  \ . \ee
Both coefficients   vanish  in the  case of  the  maximally (1,1) supersymmetric YM theory (SYM)  in 6d which can be 
obtained by dimensional 
 reduction from  the 10d  SYM  giving  $N_1=1, \ N_0=4, \ N_{1\ov 2} = 2$.\foot{Equivalently, 
 if  we   consider the  (1,0)  SYM coupled to one   adjoint hypermultiplet we  get the same  1-loop 
 finite theory  (cf. \cite{Buchbinder:2017ozh}).} 
Note  that the expression   for  the coefficient 
 $\b_3$  of the $F^3$ invariant   in \rf{83}  happens to  coincide  with the  
number of  effective  degrees of freedom  and so it vanishes also in the case of (1,0)  6d SYM  
 where $N_1=1, N_0=0, N_{1\ov 2} =1$. 
This is consistent with the fact that  the  only possible 
(1,0)  6d super-invariant   is the one   with the  bosonic   part 
containing    $ (D_m F_{mn})^2$, i.e.  the $F^3$   invariant is ruled out by (1,0) 
supersymmetry (see  \cite{Ivanov:2005qf}). 

As we shall find below,  in the  case of  the self-dual B-field  
the  divergent part of the effective action $\G_+$ in \rf{25} is given by \rf{820}, \rf{82} 
with $\beta_2 = -27, \  \beta_3 = -57$. 
In the case of   the non-chiral  $B$-field  with  the effective  action in \rf{12} 
these  coefficients are doubled,   in agreement with \rf{26}. 
Thus, in the presence of $N_T$   self-dual tensors,  the coefficients  in \rf{83}   
become
\be \la{833}
&&\b_2 = -27 N_T - 36 N_1 +   N_0  + 16 N_{1\ov 2} \ , \ \ \ \ \ \ \qquad\ \ \ \\
&&\b_3 =- 57 N_T +  4 N_1 +   N_0  -  4 N_{1\ov 2}  \ . \la{835}\ee
Here all fields  are assumed to be in the adjoint representation; otherwise  
$N_T, N_0, N_{1\ov 2} $ are to be rescaled by the corresponding factors  $T_R/C_2$.

\iffa 
We may also add (1,0) hyper multiplet:    ($N_0=4, N_{1/2} =1$)
 If we   expect   that  (1,0) tensor multiplet 
is taken in some representation $R'$  while hyper in representation $R$  
  we would then get  from the sum of (1,0)  hyper and (1,0) tensor   multiplets  coupled as ``matter"  
   to (1,0)  SYM  (adding factors of  representation indices)
\begin{align}  
&\b_2 =   C_2 (  - 36    + 16)   +  T_{R}   ( 4 + 16)  +   T_{R'} ( -27  + 1    + 16)    \ , \no  \qquad \\
&\b_3 =  C_2 (   4    -  4)  +      T_{R} ( 4 -4)  +   T_{R'}  ( - {57} +   1  -  4  )  \ . \la{102} 
\end{align}
\fi

\renewcommand{\theequation}{4.\arabic{equation}}
 \setcounter{equation}{0}
 \section{Calculation of one-loop   divergences }

Let us now  compute  the  coefficients  in the 
 logarithmically divergent part of the one-loop  effective actions $\G_+$  and $\G$  
for the self-dual  \rf{21} and the non-chiral   \rf{10} 2-form   models.
 
We adopt dimensional regularization, by continuing the theory to $\dd=6-2\varepsilon$ dimensions. 
Since the sixth direction is treated separately in the classical action and in the gauge fixing condition  ($B_{i6}=0$),  
 it is natural to keep it  one-dimensional,  while continuing the remaining 5 directions, 
setting   $ 6 = 1+5 \to  1+ d, \  d=5- 2\varepsilon$.
 Within the  dimensional regularization we consider  
 an analog of the four-dimensional helicity scheme, 
where all the  momentum numerator algebra  is first  done in an integer number of dimensions
and then the scalar   integrals  are continued  to $d$    dimensions. 
This guarantees that the number of physical states in loops is unchanged by the regulator.

To  find   the  coefficients  in  the  divergent part of the effective action 
one may   compute, e.g.,  the   terms 
quadratic and cubic in the  vector field $A$  and compare them with \eqref{82}.
This is  what we will do  below. 
Alternatively,   one may find  the  terms    with  six    powers of $A$
which appear   in \rf{82}  without derivatives and thus can be  isolated 
by taking the  non-abelian field  $A$ to be constant. This will be done in Appendix \ref{appC}   on the example 
of the self-dual model~\rf{21}.


\iffa 
We shall
 To carry out the calculation in perturbation theory we split the quadratic operator $\Delta$  in \rf{11},\rf{24} 
  as
\be
\Delta = \Delta^{(0)}+\Delta^{(1)} + \Delta^{(2)}\ ,
\label{decomposition}
\ee
where $\Delta^{(n)}$ contains $n$ powers of the  background field $A$. 
\fi 

 \subsection{Self-dual  $B$-field model}

The  effective action   corresponding to  the classical action  \rf{21}  is\foot{Compared to \rf{25},\rf{26}  here we include  the 
$A$-independent  factor $\del_6$  in the kinetic operator  making it symmetric.} 
\be \la{41}
\G_+ = \ha \log \det \Delta_+  \  , \qquad  \qquad   \Delta_+ B_{ij} =  - \del_6 \hat \OO_+  B_{ij} 
= -\del_6 ( \del_6  B_{ij}   + \te \frac{i}{2}  \epsilon_{ijkmn}  D_k  B_{mn} )  \ .
\ee
The operator $\Delta_+$ is thus linear in  the background field $A_i$, i.e.\foot{Here  $a,b,c$ are Lie algebra indices.
We assume that $B$ is in adjoint representation; otherwise  $t^a_{bc} =
-i f^{a}_{\ bc} $ is to be replaced by the corresponding  hermitian 
generators.}  
\begin{align} \la{42} 
&\Delta_+ = \Delta^{(0)}+\Delta^{(1)}\ , \qquad \\
&
[\Delta^{(0)}]^{ab}_{ij, mn} =\te   -\delta^{ab}\big(\delta_{ij, mn}\partial^2_6  +  \frac{i}{2} \epsilon_{ijkmn} \partial_6 \partial_k \big)
\ , \qquad 
[\Delta^{(1)}]^{ab}_{ij, mn} =   - \frac{i}{2} f^{acb} \epsilon_{ijkmn}  A^c_k \partial_6 \ . \la{43}
\end{align}
Expanding the non-trivial part of  $\G_+$    in powers of $A$,  we have 
\be
\te \G_+= \G_2 + \G_3 + ....\ , \qquad   &&\te \Gamma_2= - {1\over 4} \tr\big[(\Delta^{(0)})^{-1} \Delta^{(1)} (\Delta^{(0)})^{-1} \Delta^{(1)}\big] \ ,
\label{44}
\\
&&\te \Gamma_3=  {1\over 6} \tr\big[(\Delta^{(0)})^{-1} \Delta^{(1)} (\Delta^{(0)})^{-1}\Delta^{(1)} (\Delta^{(0)})^{-1}  \Delta^{(1)}\big] \ .\no 
\ee  
Since the background field $A_i$  is independent of $x_6$, the trace projects out all terms containing an odd number of $\partial_6$ factors  and  also 
produces  an overall   factor of    length     $L_6=\int dx_6$.
As was already mentioned  in section 2, 
together with the symmetry of the gauge-fixed action~\rf{21} under 
$\partial_6\rightarrow -\partial_6$ combined with 
$\epsilon_5\rightarrow -\epsilon_5$, this implies the 
effective action $\G_+$ is parity-even.

The evaluation of traces is standard, by using   momentum space basis of states and assuming that the background field  is 
$
A_i^a(x_k) = \int  {d^5s \ov (2\pi)^5 }\tA_i^a (s) e^{i s_k x_k}  . 
$
The matrix element of $(\Delta^{(0)})^{-1}$ in  momentum  representation 
is the  free $B$-field propagator 
\be
&&\langle p| (\Delta^{(0)})^{-1}|p \rangle\   \to \ \delta^{ab} \ P^{jk}_{mn}(p_i, p_6)\ , \no \\
&& P^{jk}_{mn}(p_i, p_6) \equiv {1 \over ( p_i^2 +p^2_6)} \Big( \delta^{jk}_{mn}- {\te {i\over 2} } { \eps_{mnq}^{~~~~~jk} p_q \over p_6}  
+  2 {p^{[j} p_{[m} \delta^{k]}_{n]}\over p^2_6} \Big) \ . \label{46}
\ee 
The matrix element of $\Delta^{(1)}$ is the vertex 
\be
\langle p+s| \Delta^{(1)}|p \rangle\ \to \ V_{ij}^{ab\  mn} (s_i, p_6)  \equiv  \ha f^{acb}  \eps_{ij}{}^{kmn}  p_6  \tA^c_k(s_i)  \ .
\label{477}
\ee

\subsubsection{$A^2$    term 
   \label{G2_chiral}}

Inserting complete sets of momentum eigenstates between any two operators in \eqref{44} and using  \eqref{46},\eqref{477}  and momentum  conservation, we have\footnote{The same expression may be obtained by computing the 
two-point function of $A$ and promoting it to a term in the effective action. In this approach, the numerical factors are symmetry factors and the signs related to  resummation of one-loop corrections to the $A$-field two-point function.}
\be
&&   \qquad    \G_2 =   L_6    \int  {d^5s \ov (2\pi)^5 }\  \cG_2(s) \ ,  \la{48}\\
\label{416}
&&
\   \cG_2 (s)  =  -\tfrac{1}{4}  \int \frac{dp_6 d^d p}{(2\pi)^{d+1} }    \
V_{i_1 i_2}^{cd\ j_1 j_2}(s_{ i}, p_6)\,  P_{j_1 j_2}^{k_1 k_2}(p_i, p_6)\ 
 V_{k_1 k_2}^{dc\ l_1 l_2}(-s_i , p_6)\,  P_{l_1 l_2}^{i_1 i_2}(p_i +s_i, p_6) \ ,
 \no
\ee
where  $d= 5 -2  \epsilon$  and $L_6 =\int  dx_6$. 
Since  the external field   does not depend   on $x_6$  here all the   factors have the same  6-th component  of momentum  $p_6$. 

The  background-field   gauge invariance requires  that \rf{48}  should vanish
for constant   $A_i$, i.e. for $s_i=0$. 
Setting $s_i=0$  and carrying out  index contractions  we get  
\be
\cG_2 \propto  \int \frac{{d p_6} {d^d p}}{(2\pi)^{d+1}} {3 p_i^2 +  5 p^2_6\over  (p^2_i+p^2_6)^2} 
= \frac{d-5}{d-2}\int \frac{{d p_6} {d^d p}}{(2\pi)^{d+1}} { 2\, p^2_6 \over  (p^2_i+p^2_6)^2} \ ,
\ee
where 
 we used the identity \eqref{000}. Thus, for a constant external field, the 
 $A^2$  contribution 
 vanishes  in  $d=5$  even before performing the integration over the $p_6$ momentum.

Contracting group indices   using eq.~\eqref{822},  introducing 
Feynman parameter $y$  in  the  momentum integral,  doing  tensor reduction   with the help of  \eqref{101}, 
and finally using the  identity \eqref{000}  gives the following expression for \rf{48}:
\be
\Gamma_2 &=& \tfrac{1}{4} C_2 \, L_6  \int \frac{d^5 s}{(2\pi)^5}  \,  \td A^a_i(s)\, \big(\delta_{ij}s^2- s_i s_j\big)\, \Pi(s^2)\,\td A^a_j(-s) \ , \la{410}
\\
\Pi(s^2) &=& \int_0^1 dy  \int \frac{dp_6 d^dp}{(2\pi)^{d+1}}\
\frac{(1 -y ) \big[ (1- 12 y) \, p_6^2  - 2 y\,  s^2\big]} {2 p_6^2 \big[p_i^2+p_6^2  + y(1 - y)  \, s^2\big]^2}\ . \la{411}
\ee
The $d$-dimensional integral here  is standard (cf. 
eq.~\eqref{104});   while  it is   finite  for  $d\rightarrow 5$, taking this limit before the $p_6$ integral makes the latter divergent.
To carry out the $p_6$   integral   it is convenient to  change  the variable  $p_6 \to \mu$
as  
  $p_6 = \mu \big[y(1-y) s^2\big]^{1/2}$;  then the  remaining  $\mu$-integral can be computed using 
 \eqref{001}. 
 As a result, we find  that the  divergent part of the $A^2$ term in the  effective action is
\be
\label{412}
\Gamma_{2\, \infty}=  {1\over d-5}\,   {9\, C_2\over  \ 5 \times  2^8  \pi^3 }   L_6  \int {d^5 s \over (2 \pi)^5}~    \td A^a_i(s)\, 
 s^2 \big( s^i s^j   - \delta^{ij}s^2\big)  \td A^a_j(-s) \ .
\ee  
Comparing this   with the first term   in  \rf{82},\rf{822}  
(with ${1\over d-5}$   identified  with  $- \log \Lambda$ in \rf{820})
we conclude  that   (cf. \rf{833}) 
\be \la{414}
\beta_2 =  -27 \ . 
\ee 
As was already mentioned  below \rf{823}, in the case of the $B$-field being in  generic 
representation $R$   the  coefficient  $C_2$ is to be replaced by the corresponding index $T_R$. 

\iffa
The result implies that the two-field UV-divergent part  of the effective action written in coordinate representation is
\be
\label{DFDF0}
\Gamma_2=  {9\, T_R\over 2^8 \times 5 \pi^3 }  \int d^6x  (D_i F^a_{ij})  (D_k F^a_{kj}) \log \Lambda
\qquad
\Tr[T^a_R T^b_R] = T(R)\delta^{ab}
\ .
\ee 
Here we generalized to $B$ fields in an arbitrary representation $R$; if $R$ is 
the adjoint representation $T(R)=C_2$.  We have identified that ${1\over d-5}\to - \log \Lambda$   where $\Lambda \to \infty$ is UV cutoff.  
Comparing \eqref{DFDF0} with \eqref{820} and \eqref{82} it follows that
as mentioned in eq.~\eqref{833}. 
\fi

\subsubsection{$A^3$ term}

To find $\beta_3$ in \rf{82} we need to  compute the $A^3$  part of the effective action.
The evaluation of $\Gamma_3$ in \eqref{44} follows the same  steps as that of $\Gamma_2$. 
For a $B$-field in an arbitrary representation  \eqref{44} becomes
\be
&&     \G_3 =   L_6    \int {d^5s_1 \ov (2\pi)^5 } {d^5s_2 \ov (2\pi)^5 }{d^5s_3\ov (2\pi)^5 } 
\  \cG_3(s_1, s_2,s_3) \, \delta^{(5)} (s_1 + s_2 + s_3) \  ,  \la{415}\\
\label{416}
&&  \cG_3 =\tfrac{1}{6} \int \frac{dp_6 d^d p}{(2\pi)^{d+1}}  \;\tr\Big[
                       V_{j_5 j_6}^{i_1 i_2}(s_{1i}, p_6)\,  
                        P_{i_1 i_2}^{j_1 j_2}(p_i, p_6 ) \, 
                      V_{j_1 j_2}^{i_3 i_4} (s_{2i}, p_6)
                      \cr
                && \qquad \qquad \qquad \qquad 
               \times\,   P_{i_3 i_4}^{j_3 j_4}(p_i +s_{2i}, p_6) 
                       V_{j_3 j_4}^{i_5 i_6}(s_{3i}, p_6)\,   P_{i_5 i_6}^{j_5 j_6}(p_i+s_{2i} + s_{3i} , p_6)\Big] \ .  
\ee
    Here $V^{ij}_{mn}$ is the vertex in \rf{477}   with  $f^{acb}$   replaced by  $ -i t^c$ 
 where $t^c$   is hermitian    generator   in some representation  $R$  (coming from the covariant derivative $D_i B = \del_i B - i  t^a A^a_i B$).
     To compute the  trace over  the group indices we use that 
\be
\tr(t^a\, t^b\, t^c) =\tfrac{1}{2}T_R\,  f^{abc} + \tfrac{1}{2}{\rm A}_R\,  d^{abc} \ ,
\label{groupth_3}
\ee
where ${\rm A}_R$ is the anomaly coefficient of a given representation. In adjoint representation $T_R= C_2, \ {\rm A}_R=0$. 
The momentum-dependent coefficient of the  symmetric $  d^{abc}$    tensor   part   is P-odd (containing 
one power of $\ep_5$)     and should thus vanish identically as discussed above.

After carrying out the index contraction, Feynman parametrization and momentum integration,
the divergent part of  $\cG_3$  may be written  as (in the adjoint representation)
\be
\label{A3_eff_act}
\cG_{3\, \infty}  = \frac{1}{d-5}\, 
  \frac{i}{ 15 \times 2^8 \pi^3} \,   C_2  \,  f^{a_1 a_2 a_3}   \; K^{a_1 a_2 a_3}(s_1, s_2, s_3) \ ,
\ee
 where $K^{a_1 a_2 a_3}$ is a  5d invariant
 constructed from  3   powers of the background field $\td A_i (s_k)$  and  the corresponding momenta ($s^2_r \equiv  s_r \cdot s_r$):
\be
&&K^{a_1 a_2 a_3}= 
\tA^{a_1}(s_1)\cdot s_1\  \Big[ -9 \tA^{a_2}(s_2)\cdot s_1\    \big(2 \tA^{a_3}(s_3)\cdot s_1 + \tA^{a_3}(s_3)\cdot s_3\big) \cr  
&& \ \ \ \qquad \qquad\qquad \qquad\ \ \ \ 
+ \tA^{a_2}(s_2)\cdot s_3 \ \big(\tA^{a_3}(s_3)\cdot s_1 + 9 \tA^{a_3}(s_3)\cdot s_3\big)\Big]  
    \cr
    &&  
    +  \tA^{a_1}(s_1)\cdot s_3 \  \Big[ -\tA^{a_2}(s_2)\cdot s_1\   \big(19 \tA^{a_3}(s_3)\cdot s_1 + \tA^{a_3}(s_3)\cdot s_3\big) 
    \la{333}\\
    && 
    \qquad \qquad \qquad \ \ \  + \tA^{a_2}(s_2)\cdot s_3\  \big(19 \tA^{a_3}(s_3)\cdot s_1 + 18 \tA^{a_3}(s_3)\cdot s_3\big)\Big]
    \cr
    && + \tA^{a_1}(s_1)\cdot \tA^{a_2}(s_2)\  \Big[   
    \tA^{a_3}(s_3)\cdot s_1 \ (36 s_1^2 + 36 s_2^2 - 2 s_3^2)  
      + \tA^{a_3}(s_3)\cdot s_3 \  (17 s_1^2 + 19 s_2^2 - s_3^2)\Big]
      \cr
      &&  + 
 \tA^{a_1}(s_1)\cdot \tA^{a_3}(s_3) \   \Big[  \tA^{a_2}(s_2)\cdot s_1\ (-19 s_1^2 + s_2^2 - 17 s_3^2) 
      + 
      \tA^{a_2}(s_2)\cdot s_3 \ (17 s_1^2 - s_2^2 + 19 s_3^2) \Big]
      \cr
      &&\   + 
 \tA^{a_2}(s_2)\cdot \tA^{a_3}(s_3) \Big[ \tA^{a_1}(s_1)\cdot s_1\  (s_1^2 - 19 s_2^2 - 17 s_3^2) 
      +  2 \tA^{a_1}(s_1)\cdot s_3\  (s_1^2 - 18 s_2^2 -18  s_3^2)\Big]
      . \no  
\ee
It simplifies   in the   transverse  background  gauge  $s_i \tA^a_i (s) =0$:  
\be K^{a_1 a_2 a_3} &=& -19\,  s_3\cdot \tA^{a_1}(s_1)\ s_1\cdot \tA^{a_3}(s_3)\ (s_1-s_3)\cdot \tA^{a_2}(s_2)
&\cr&&
+ \big[ 18(s_1^2+s_2^2)-s_3^2\big] \  \tA^{a_1}(s_1)\cdot \tA^{a_2}(s_2) \  s_1\cdot \tA^{a_3}(s_3) 
\cr
&&+ \big[18(s_2^2+s_3^2)-s_1^2\big] \ \tA^{a_2}(s_2)\cdot \tA^{a_3}(s_3) \  s_2\cdot \tA^{a_1}(s_3) \ 
\cr&&
+ \big[18(s_3^2+s_1^2)-s_2^2\big] \  \tA^{a_3}(s_3)\cdot \tA^{a_1}(s_1) \  s_3\cdot \tA^{a_2}(s_2)   \ .
\la{4118}\ee
Comparing this to the   two terms in \rf{82} (which both contribute to  the $A^3$  term)
and using that $\b_2$ was already determined   in \rf{414}  we conclude that (cf. \rf{835})
\be
\beta_3 = - 57 \ .\la{419}
\ee  
We have confirmed this result independently   by computing  constant  $A^6$ term in  the effective action   in Appendix \ref{appC}.

\subsection{Non-chiral  $B$-field model}

 In the  non-chiral theory   \rf{10},\rf{11}   the effective action  \rf{12} is   given by 
\be\la{421}
\Gamma = \textstyle{\frac{1}{2}} \ln\det\Delta \ , \qquad 
\qquad
\Delta B_{ij} = -(\del_6^2 + D^2)B_{ij}  + 2 \delta^{[m}_{[{i}} D^{n]} D_{j]} B_{mn} \ .
\ee
Here $\Delta$ is  quadratic in the background field $A_i$, i.e. (cf. \rf{42},\rf{43})
\be
\Delta &=& \Delta^{(0)} + \Delta^{(1)} + \Delta^{(2)}\ , \cr
[\Delta^{(0)}]_{ij,mn}^{ac} &=& \delta^{ac} \big[-\delta_{ij, mn} 
( \partial^2_i +\d^2_6  )  + 2  \delta_{[m[{i}} \del_{j]} \del_{n]} \big]  \ , 
\cr
[\Delta^{(1)}]_{ij,mn}^{ac} &=&  f^{abc} \big[  -\delta_{ij,mn} ( \partial_k A^b_k +2  A^b_k \partial_k  )
 +2 \delta_{[i[m} (A^{b}_{  n]} \partial_{j ]}   + \partial_{n]}  A^b_{j]}  + A^b_{j ]} \partial_{n]}  ) \big]\ , 
\cr
[\Delta^{(2)}]_{ij,mn}^{ac} &=&   f^{ade} f^{ebc} \big[ -\delta_{ij, mn} 
A^d_k A^b_k   + 2   \delta_{[i[m} A^d_{n]} A^b_{j]} \big] \ .  \la{422} 
\ee
The quadratic and cubic  in $A_i$  parts of the effective action   have the structure (cf. \rf{44})
\be
&&\G= \G_2+ \G_3 + ... \ , \qquad 
\Gamma_{2} =  \tfrac{1}{2} \tr \big[( \Delta^{(0)})^{-1} \Delta^{(2)} \big]
  - \tfrac{1}{4} \tr \big[(\Delta^{(0)})^{-1} \Delta^{(1)}(\Delta^{(0)})^{-1} 
  \Delta^{(1)}  \big] , \ \ \qquad \label{423}
\\
&&\Gamma_{3} = -  \tfrac{1}{2} \tr \big[( \Delta^{(0)})^{-1} \Delta^{(2)}
(\Delta^{(0)})^{-1} \Delta^{(1)} \big]
  +  \tfrac{1}{6} \tr \big[   (\Delta^{(0)})^{-1} \Delta^{(1)}(\Delta^{(0)})^{-1} 
  \Delta^{(1)} ( \Delta^{(0)})^{-1} \Delta^{(1)}\big] . \no
\ee
The analogs of the relations \rf{46},\rf{477}  in momentum representation  are 
\be
&&\langle p| (\Delta^{(0)})^{-1}|p \rangle   \to   \delta^{ab}  P^{ij}_{mn}(p_k, p_6) \ , \qquad 
P^{ij}_{mn}(p_k, p_6)=  {1\over ( p_i^2 +p^2_6)} 
\Big( \delta^{ij}_{mn}+  2 {p^{[i} p_{[m} \delta^{j]}_{n]}\over p^2_6} \Big)\ , 
\la{4233}  \\
&&\langle p+s| {\Delta^{(1)}} |p\rangle \to   V^{(1)}{}^{ab \ mn}_{ij} (p_k, s_k) \ \cr
 && 
\qquad \qquad \qquad \qquad= -i f^{acb} \Big[\delta^{mn}_{ij} \tA^c_k  \big( s_k +2  p_k\big)
+2 \delta^{[m}_{[j} \big(  \tA^c_{i]} s^{n]} +\tA^{n]c} p_{i]}  + \tA^c_{i]}  p^{n]} \big)\Big]\ , \la{424}
\\
&&\langle p+s_1+s_2| {\Delta^{(2)}}|p \rangle  \to   V^{(2)}{}^{ab \ mn}_{ij}(p_k, s_{1k},s_{2k})
= f^{ade} f^{bce}  \Big(\delta^{mn}_{ij}  \tA^d_k \tA^c_k + 
{2} \delta^{[m}_{[j}  \tA^{n]d} \tA^c_{i]} \Big) .~~\qquad \la{425}
\ee

\subsubsection{$A^2$ term}

The first term in $\G_2$   in \rf{423}  is a tadpole   which vanishes in dimensional regularization;
the  second term gives (using the same notation as in  \rf{416})
\be
   \qquad    \G_2 =   L_6    \int  {d^5s \ov (2\pi)^5 }\  \cG_2(s) \ , 
\label{4166}\ee
$$
\   \cG_2  =  -\tfrac{1}{4}  \int \frac{dp_6 d^d p}{(2\pi)^{d+1} }    \
V^{(1)}{}_{i_1 i_2}^{cd\ j_1 j_2}(s_{ i}, p_6)\,  P_{j_1 j_2}^{k_1 k_2}(p_i, p_6)\ 
 V^{(1)}{}_{k_1 k_2}^{dc\ l_1 l_2}(p_i+s_i,-s_i)\,  P_{l_1 l_2}^{i_1 i_2}(p_i +s_i, p_6).
$$
Following similar    steps as in the  self-dual model  in  section~\ref{G2_chiral}  
we find  ($p^2= p_i p_i$)
\be\la{427}
 \cG_2
= -\tfrac{3}{2} C_2  \int_0^1 dy  \int \frac{dp_6 d^d p}{(2\pi)^{d+1} }   \,  Q (s_i, p_k, p_6, y) \ , 
\ee
$$Q=
\Big(    \big[\tfrac{1}{2} -   y (1 - y)\big] s^2   +  y^2 (1 - y)^2  \frac{s^4}{p_6^2} 
 + \tfrac{8}{5} p^2  + \big[5 - 26 y (1 - y) \big] \frac{s^2 p^2}{10 p_6^2} 
   + \frac{12 p^4}{5 p_6^2} \Big) \tA^a(s)\cdot \tA^a(-s)
   $$
\be    
 -
\Big[\tfrac{1}{2} -3 y (1 - y)  - y^2 (1 - y)^2 \frac{s^2}{p_6^2} - \big[1 - 18 y(1 - y) \big] \frac{p^2}{10 p_6^2}   \Big] 
\ s\cdot \tA^a(s) \  s\cdot \tA^a(-s) .   
\la{428}
\ee
It is useful to use eq.~\eqref{000}  to relate the integrals with loop momenta in the numerators to scalar bubble integrals.
Unlike the self-dual  theory case,   here the $A^2$ term  takes  a   gauge-invariant   form  only after 
 one carries out all the  integrals;  its divergent  part is  found to be 
 \be
\label{429}
\Gamma_{2\, \infty}=  {1\over d-5}\,   {9\, C_2\over  \ 5 \times  2^8  \pi^3 }   L_6  \int {d^5 s \over (2 \pi)^5}~    \td A^a_i(s)\, 
 s^2 \big( s^i s^j   - \delta^{ij}s^2\big)  \td A^a_j(-s) \ .
\ee  
 This is twice   the value in the self-dual case   \rf{412}, i.e.   the corresponding  $\beta_2$ coefficient in \rf{82} 
 is  (cf.  \rf{414}) 
\be\la{430}
\beta_2 = - 54 = 2 \beta_2^\text{self-dual} \ .
\ee

\subsubsection{$A^3$ term}

Unlike the case of  $\Gamma_2$, 
   the matrix element of $\Delta^{(2)}$  in \rf{425}  contributes nontrivially  to $\Gamma_3$ 
in \rf{423} (cf. \rf{415})
\be
&&     \G_3 =   L_6    \int {d^5s_1 \ov (2\pi)^5 } {d^5s_2 \ov (2\pi)^5 }{d^5s_3\ov (2\pi)^5 } 
\  \cG_3(s_1, s_2,s_3) \, \delta^{(5)} (s_1 + s_2 + s_3) \  ,  \la{4155}\\
\label{4167}
&&  \cG_3 =\int \frac{dp_6 d^d p}{(2\pi)^{d+1}}  \;\Big[
- \tfrac{1}{2} 
V^{(2)}{}_{i_1 i_2}^{cd \ j_1 j_2}(p_i,  s_{1i}, s_{2i}) \,  P_{j_1 j_2}^{k_1 k_2}(p_i, p_6) \, 
 V^{(1)}{}_{k_1 k_2}^{dc\ l_1 l_2}   (q_i 
  ,  s_{3 i}) \,  P_{l_1 l_2}^{i_1 i_2} (q_i 
  , p_6)
 \Big|_{q = p- s_{3}} \no \\
&&\qquad \qquad \qquad \qquad  \ +    \tfrac{1}{6}    V^{(1)}{}_{j_5 j_6}^{de\ i_1 i_2}(p_i, s_{2,i})\,  
                        P_{i_1 i_2}^{j_1 j_2}(p_i, p_6 ) \, 
                      V^{(1)}{}_{ j_1 j_2}^{ef\ i_3 i_4} (q_i, s_{1i})
                      \cr                && \qquad \qquad \qquad  \qquad  \times\,  
               \,  P_{i_3 i_4}^{j_3 j_4}(q_i , p_6) \, 
                       V^{(1)}{}_{j_3 j_4}^{fd\ i_5 i_6}   (r_i, s_{3i})      \,   P_{i_5 i_6}^{j_5 j_6}(r_i , p_6) \Big|_{q=p-s_1,\,  r=p-s_1-s_3}
                       \Big]
                        \ .  \ee
Here we used   \rf{groupth_3}  to do the group index contraction 
(with the momentum-dependent coefficient of $d^{abc}$  again  vanishing in general)
and considered the adjoint representation.  Introducing Feynman parameters
and shifting the integration variable in such
a way  that the denominator becomes a symmetric function, we  
use  the $SO(d)$ symmetry to express the tensor  momentum integrals in terms of the scalar  ones. 
Further using eq.~\eqref{000} the integrals with various powers of the $d$-momentum  can be 
 reduced to  scalar triangle and/or bubble integrals. Evaluating first the $d$-dimensional integral
  and then appropriately changing the variable  of  the $p_6$ integral one   can 
  decouple  the  latter  from that over the Feynman parameters. 
As before, the   logarithmic UV divergence we are interested in emerges after the last 
($p_6$) integral is evaluated using \rf{001}.

The contribution of the 
 first   structure  in   \rf{4167}  written in coordinate representation gives  
a term proportional to $  {1\ov d-5}   L_6 \int d^5 x \   f^{abc} A_i^a A_j^b  \del^2  (\partial_i A^c_j - \partial_j A^c_i)  $
which  matches  the cubic term in $\tr (\partial_m F_{mn})^2$  in \rf{82}. 
Gauge invariance and consistency with 
the $A^2$ term  calculation \rf{429}  are restored by the inclusion of the second term in \rf{4167}. 

The  full computation
is straightforward but tedious  so we simply state  that 
the final result is consistent with \eqref{82} with $\beta_2$ 
found above in \rf{430}   and with $\beta_3$     being   again  twice the value  in the self-dual case  \rf{419}: 
\be
\beta_3 = - 114 = 2 \beta_3^\text{self-dual}  \ .\la{431} 
\ee

\renewcommand{\theequation}{5.\arabic{equation}}
 \setcounter{equation}{0}
 \section{Concluding   remarks } 

In this paper we have  studied a   model   of  6d 2-form $B$-fields    in some
  representation of an  internal symmetry  group $G$    coupled  consistently  to a non-abelian  gauge field  $A$
which ``lives"  only in a 5d  subspace.
We computed   one-loop  logarithmic divergences in such a theory by integrating out the $B$-field and  
   treating the gauge field $A$ as  a background.  The resulting  divergent part of the effective 
action \rf{820},\rf{82}   contains the terms $\sim \tr [3\b_2 (D_mF_{mn})^2  - {2}\b_3  F_{mn} F_{nk} F_{km} ]$  
  with the coefficients $\b_2,\b_3$  given by 
\rf{833},\rf{835} or \rf{414},\rf{419} in the case  of  the 
self-dual theory (and twice these values  in the 
case of  the   non-chiral $B$-field model given in~\rf{430},\rf{431}).

The presence of these divergences   suggests  that   in the   full theory where  $A$-field (or its 6d extension) 
 is also quantized 
 the   higher-derivative $c_1 (DF)^2  + c_2 F^3$   terms   should be added   
 to the bare  6d action.\foot{$(DF)^2  +  F^3$  theory is of course classically   conformally invariant 
 but this   symmetry  will be anomalous  at loop level.
 Let
 us note that  the presence of similar 
$(DF)^2$   terms in 6d theory was  suggested    by requiring  dual conformal symmetry in six dimensions in 
\cite{Bhattacharya:2016ydi}, though precise connection of this 
 to the present work is not clear to us.}
   One   may  hope that these  divergences 
 could  be cancelled   by adding  some  other fields  to the model   and imposing, e.g., supersymmetry constraint. 
 For a collection of $N_T$    self-dual tensors, $N_1$   standard 2-derivative
 YM  vectors, $N_0$   real scalars   and $N_{1\ov 2}$ Weyl 
 fermions in 6d  coupled to background vector, we conclude that\foot{Note that quantizing the 5d gauge field with the action $L_6 \int d^5x [ c_1  (D_iF_{ij})^2  +  c_2 F_{ij}^3]$
  will not produce extra  one-loop  logarithmic  UV divergences as this theory is effectively  5-dimensional  one. 
  It would   be interesting   to add  also the one-loop contribution of  
the genuine 6d  non-abelian vector  model  
 with the classical action $\int d^6x \, \tr [ c_1  (D_\m F_{\m\n})^2  +  c_2 F_{\m\n}F_{\n\l} F_{\l\m}]$. 
  Choosing the {\it background}  gauge field  to be a   5d one   we would then get     additional 
  contributions to $\b_2$  and $\b_3$. 
 }
 \be \la{839}
\b_2 = -27 N_T - 36 N_1 +   N_0  + 16 N_{1\ov 2} \ , \ \ \ \ \ \ \qquad\ \ \ \b_3 =- 57 N_T +  4 N_1 +   N_0  -  4 N_{1\ov 2}  \ .
\ee
Thus  the  self-dual $B$-field  coupled minimally to a 5d gauge  field 
contributes  to  the $\b_2$  in the logarithmic divergent part of the   effective action 
with the  same  sign  as  the  YM 
vector.   A  somewhat  unexpected feature is that  its  contribution to $\b_3$   turns out to be  opposite 
in sign compared to  other standard 2-derivative bosonic  fields. 
A naive  expectation   could be  that  each field should contribute to $\b_3$ 
 proportionally  to its  number of  dynamical degrees of  freedom: 
\be \la{888}   
\nu = 3 N_T +  4 N_1 +   N_0  -  4 N_{1\ov 2} \ . 
\ee
One may  formally consider    fields    that form  6d supermultiplets 
containing  self-dual $B$-field 
 and  couple them  to a background   5d gauge field. 
In the case of   (1,0) tensor multiplet    with   $N_T=1, \ N_1=0, \ N_0= 1, \ N_{1\ov 2}= 1$ 
a natural  coupling is  to (1,0) SYM ($N_1=1, N_0=0, N_{1\ov 2} =1$);  
  in this case we   would  expect to get  $\b_3=0 $
as the $F^3$-invariant should be prohibited by supersymmetry. 
However,   while  $\nu$ in \rf{888} indeed    vanishes in this case, from \rf{839}  we get  $\b_3 =-60= 2\b_2$. 
This suggests that the model considered in this paper 
(with $A_i$ treated as a 5d background field) 
 does not admit a (1,0) supersymmetric extension. 
This may not be surprising as the model lacks 6d Lorentz symmetry.\foot{If  we naively  consider the case of (2,0) tensor multiplet with  $N_T=1, \ N_1=0, \ N_0= 5, \ N_{1\ov 2}= 2$, we obtain  $\beta_{2}=-{1\over 6} \b_3=10$.
 It  is also interesting to note that    similar  $\int d^6x \, \tr [ c_1  (D_\m F_{\m\n})^2  +  c_2 F_{\m\n}F_{\n\l} F_{\l\m}]$
 divergences  (or contributions to stress tensor  anomaly)  
 appear if one couples  (2,0)   tensor multiplet 
 to the R-symmetry $SO(5)$     vector  gauge   field \cite{Manvelyan:2000ea};  as  
   the $B$-field   is singlet    under the $SO(5)$,  there the contribution  comes only   from  the coupling 
   of the $SO(5)$  field to  the scalars and   fermions. 
  }

A possible role or application of the  non-abelian 
 coupled $(B,A)$ model   discussed in this paper  remains an open question. 
  It might  be related to some  intersecting brane configuration 
 where a 5d  gauge field   lives  on a 5d  brane  ``defect".
  Another  option is that the 5d $A$-field   may  happen to play   an   auxiliary 
 role and    eliminating it one may  end up with an effective  interacting theory of a set of $B$-field only. 
 Yet another   possibility  is the existence of a 
    generalization where   the $A$   gauge field becomes fully  6-dimensional, 
 Lorentz invariance is   formally restored  but the resulting  classical action might become effectively non-local.

 \iffa
An important   open question  is about  existence of  a  consistent theory    where $A$-field 
is  fully  6-dimensional 
(i.e.  with  $A_6\neq 0$, $\partial_6 A_{\mu}\neq 0$)  and thus 6d Lorentz invariance is restored. 
It would   also  be interesting to understand  possible  meaning  of $A$-field  in the context of    connection 
of self-dual  $B$-field to (2,0)  theory  where  an independent vector is not expected 
 but   might be effectively   related to self-interactions  of $B$-field. 
\fi


\subsection*{Acknowledgments}
KWH would like to thank P.-M. Ho and E. Witten for discussions of the non-local gauge symmetry and remarks on M5-branes at an early stage of this work.  
AAT is grateful to R. Metsaev and K. Mkrtchyan for related discussions  of 6d theories. 
The work of KWH was supported
in part by the National Science Foundation under
Grant No. PHY-1620628. 
The work of RR was supported by the Department of Energy under Award Number DE-SC0013699.
AAT  was  supported by the STFC grant ST/P000762/1  
and  by  the Russian Science Foundation grant 14-42-00047 at Lebedev Institute.

\appendix
\renewcommand{\theequation}{A.\arabic{equation}}
\section{Free  partition function \la{appA}}
\setcounter{equation}{0}

For  a  free  rank 2  antisymmetric  tensor   $B_{\m\n}$   in $d$ dimensions 
 with action $\int d^d x H_{\m\n\l} H^{\m\n\l},  H= d B $,
 the partition function in the covariant Feynman-like gauge  (found   by 
  adding  the $(\del_\m B_{\m\n})^2 $  term to the action) 
  is \ci{Schwarz:1979ae}
\be \la{a1}
Z=\Big[  { (\det   \Delta_1 )^2  \ov \det \Delta_2 \  (\det \Delta_0)^3    } \Big]^{1/2} \ , 
\ee
where  the free  Laplacians $\Delta_n = - \del^2$   are defined 
on rank $n$  antisymmetric tensors. 
The number of dynamical degrees of freedom $\nu_2(d) $   of  
rank 2  tensor in $d$ dimensions 
extracted from  the representation of 
  $Z$ as  $ [\det \Delta_0]^{- \nu/2}$
is then 
\be \la{a2} 
\nu_2(d) =  C^{2}_{d-2}  = \ha (d-2) (d-3) \ , \qquad \qquad  \nu_2(6) = 6\ .  \ee 
For   a  self-dual tensor in 6 dimensions we   should  get  $\nu_{2,+}(6)= 3$.  
Eq.\rf{a1}   may also  be  written as 
\be \la{a3}
Z=\Big[  { \det \Delta_{1\perp}   \ov \det \Delta_{2 \perp}  \det \Delta_0  } \Big]^{1/2} \ , 
\ee
where $\Delta_{n\perp}$   are defined on transverse tensors.\foot{Note that    $\det \Delta_1=  \Delta_{1\perp}  \det \Delta_0, \ \ 
   \det \Delta_2=  \det \Delta_{2\perp} \det   \Delta_{1\perp} $.}
        The count of degrees of freedom in 6d  goes as  follows:  $\nu_2(6)=  (15-5)  +1   - (6 -1) =6$   ($\del_\m B_{\m\n}=0$ gives $ 6 - 1=5$ conditions, etc.). 

The equivalent results are found   also   in the  ``axial" gauge  $B_{i6}=0$  ($i=1, ..., 5$)  where  $H_{6ij} = \del_6 B_{ij}, \ \ H_{ijk} =3 \del_{[i} B_{jk]}$. Separating the 5d transverse part  as 
$
B_{ij} = B^\perp_{ij} + \del_i b_j - \del_j b_i  $
and integrating over $b_i$ one finds that   the resulting determinant
cancels against the  ghost and Jacobian factors   and    we end  up  with 
\be \la{a5} 
Z=  { 1   \ov \big[\det  \Delta_{ \perp}  \big]^{1/2} }\ , 
\ee 
where $ \Delta_{ \perp}$ is  the  6d Laplacian  defined on $B^\perp_{ij} $.
Thus \rf{a5}   describes  $\ha\times  4 \times 5  - (5-1) = 6$    degrees of freedom
(for similar discussion in the 4d  vector case  see  eqs. (2.14),(2.15) in \ci{Beccaria:2014jxa}).

Analogous    considerations in  the   self-dual case   described by the classical actions 
 \rf{15} or \rf{18}  lead to the partition function  \rf{17}  or  the ``square root" of 
 \rf{a5}  (cf. \rf{20}). 

\renewcommand{\theequation}{B.\arabic{equation}}
\setcounter{equation}{0}
\section{Useful integrals  
 \la{appB}}

We use  the following   standard  integrals:
\be
\int \frac{d^d q}{(2\pi)^d}  \frac{1}{(q^2 + X)^m} &=& \frac{1}{(4\pi)^{d/2}}\frac{\Gamma(m-d/2)}{\Gamma(m)} \frac{1}{X^{m-d/2 }}
\label{104}\ , \\
\label{101}
\int \frac{d^d q}{(2\pi)^d} \frac{q^i q^j}{(q^2+X)^n} &= &\frac{1}{d} \int \frac{d^d q}{(2\pi)^d} \frac{q^2\delta^{ij}}{(q^2+X)^n}\ ,  \\
\int \frac{d^d q}{(2\pi)^d} \frac{q^i q^jq^k q^l}{(q^2+X)^n} &=& \frac{1}{d(d+2)} \int \frac{d^d q}{(2\pi)^d} 
\frac{(q^2)^2(\delta^{ij}\delta^{kl}+\delta^{ik}\delta^{lj}+\delta^{il}\delta^{jk})}{(q^2+X)^n}\ ,
\ee
 and  the  identity
\be
\label{000}
\int \frac{d^d q}{(2\pi)^d}   \frac{q^{2n}}{(q^2 + X)^m} 
= -\frac{d+2(n-1)}{d+2(n-1) -2(m-1)}\, X\, \int \frac{d^d q}{(2\pi)^d}  \frac{q^{2n-2}}{(q^2 + X)^m} \ .
\ee 
The     integral   used  for evaluation of one-dimensional  integrals  over $p_6$   is 
\be
\int_{-\infty}^{+\infty} d\mu\ \frac{\mu^{2n}}{(1+\mu^2)^{m-d/2}} = \frac{\Gamma\big(m-n-(d+1)/2\big)\Gamma(n+1/2)}{\Gamma(m-d/2)} \ , 
\qquad
m,n\in {\bf Z}
\ .
\label{001}
\ee

\renewcommand{\theequation}{C.\arabic{equation}}
\setcounter{equation}{0}
\section{$A^6$ term in  effective action of  self-dual B-field\la{appC}}
To  obtain  the $\beta_2$ and $\beta_3$ coefficients in the divergent part of the effective action \rf{82} 
one may either compute  the derivative-dependent $A^2$ and $A^3$   terms as was done in section 4, 
or consider a constant  non-abelian  $A^a_i$-field   and  find  the  coefficients of the  independent $A^6$ terms  in $\G$. 
For constant  potential one  has $F^a_{ij} = f^{abc} A^b_i A^c_j $  and thus 
\be\la{c1} 
D_j F_{ij}^a \, D_k F_{ik}^a&=& f^{abc} f^{cde} f^{afh} f^{hgw}  A^b_j A^d_i A^e_j   A^f_k A^g_i A^w_k \ , \qquad \cr
f^{adg} \, F^a_{ij} F^d_{jk} F^g_{ki}&=& f^{adg} f^{abc} f^{def} f^{ghw}  A^b_i A^c_j   A^e_j A^f_k   A^h_k A^w_i\ .
\ee
 It is sufficient to consider the $SU(2)$ case where  $f_{abc}=\eps_{abc}$  ($a=1,2,3$). 
The  effective action  corresponding to the self-dual model \rf{41},\rf{42},\rf{43}  in a  constant  
non-abelian  $SU(2)$  background  potential  $A^a_i$   may be written as 
\be\la{c3}
 \Gamma_+  = {\te{1\over 2}} \int d^6x  \int {d^6 p \over (2 \pi)^6} \tr \ln \big[1  +  (\Delta^{(0)}(p))^{-1}  \Delta^{(1)}(p) \big]\ , 
\ee 
where  the propagator $(\Delta^{(0)})^{-1} $ and  the vertex 
$\Delta^{(1)}$  (linear in $A$)   are given by  \rf{46},\rf{477}  in momentum representation. 
As a result, the $A^6$   structures coming out of  the terms  $\sim (P_{ij}^{mn}\eps_{mnr k l} A^a_{r} p_6)^6$
are   given by 
\be
\label{c33}
\Gamma_6&=& \int d^6x   \int {d p_6 \ov 2 \pi } {  d^d p \over (2 \pi)^d}\  {1\over (p^2+ p_6^2)^6\,   p_6^2 }
{\te  \Big[ {1\ov 35} \big(\cTheta_1+{3} \cTheta_2-{4} \cTheta_3\big) }{p^8} \nn\\
&&\te\qquad \qquad  -  {1\ov 3} \big({221\over 210} \cTheta_1-{47\over 14} \cTheta_2+{7\over 3} \cTheta_3 \big)  {p^8}
+ \big({3\over 10} \cTheta_1+{9\over 14} \cTheta_2-{36\over 35} \cTheta_3\big) {p^4 p^4_6} \nn\\
&&\te \qquad \qquad  +  \big({7\over 10} \cTheta_1 -{3\over 2} \cTheta_2 +{1\over 5} \cTheta_3 \big) {p^2 p^6_6}
+  \big({5\over 6} \cTheta_1-{1\over 2} \cTheta_2+{2\over 3} \cTheta_3\big) {p^8_6}\Big] \ ,
\ee   
where $p^2= p_i p_i$   and  the 6-volume factorizes.
We  introduced  the following notations for the $A^6$ contractions  
\be\la{c2} 
\cTheta_1 = Q^{ac} Q^{ab} Q^{bc} \ , \qquad
\cTheta_2 =  Q^{aa} Q^{bc} Q^{bc} \ ,\qquad 
\cTheta_3 =  Q^{aa} Q^{bb} Q^{cc} \ , \qquad Q^{ab}\equiv  A^a_i A^b_i \ , 
\ee 
in terms of which the two invariants in \rf{c1} have the forms 
\be\la{c334}
F^3  \equiv f^{adg}  F^a_{ij} F^d_{jk} F^g_{ki} 
= 2 \cTheta_1-3 \cTheta_2+\cTheta_3 \ ,  \qquad (DF)^2  \equiv  D_j F_{ij}^a \, D_k F_{ik}^a
= \cTheta_1-2 \cTheta_2+\cTheta_3 \ .
\ee  
Using  \rf{000}
we can rewrite  the integrals in \rf{c33}  as  (setting $d=5$ in the coefficients) 
\be\la{c4}
\int {d^d p} {1\over (p^2+ p_6^2)^6\, p^2_6 } \big\{{p^8},~{p^8},~{p^4 p^4_6},~{p^2 p^6_6}\big\}
= \big\{-231, 21, \tfrac{7}{ 3}, 1 \big\} \int {d^d p} { p^6_6 \over (p^2+ p_6^2)^6}\ .
\ee 
Then   the effective action \eqref{c33} takes   the form consistent  with  gauge invariance  (cf. \rf{c334})
\be
&&\Gamma_6= -{\te {16 \over 15}}  c  \int d^6x  \big(11 \cTheta_1- 3 \cTheta_2- 8 \cTheta_3\big)  
=  {\te{ 16 \over 15}}  c   \int d^6x \big[ 27 (DF)^2-19 F^3 \big]   \ ,\la{c5} \\
&&\qquad \qquad  c\equiv  \int {d p_6 \ov 2 \pi } {  d^d p \over (2 \pi)^d  } {p^6_6  \over (p^2+ p_6^2)^6} 
 = -  {1 \ov 2^{11}  \pi^3  (d-5)}   + ... \ . \la{c6}  \ee 
Here  to isolate the  UV logarithmic  divergence 
we  integrated over $p_6$, used \rf{104}  and  took  $d \to 5$  ignoring IR singularity 
(which is related to the expansion in powers of constant $A$)
\be 
\la{c7}
 \int^\infty_{-\infty}  {d p_6 \ov 2 \pi }  {p^6_6  \over (p^2+ p_6^2)^6} = { 3 \ov 512   p^5} \ , \ \ \ \ \ 
\int  {  d^d p \over (2 \pi)^d  }  { 1 \ov p^5} \to  { 1 \ov (4 \pi)^{d/2} } { \Gamma( {5-d \ov 2}) \ov \Gamma( { 5\ov 2}) }
=-  {1 \ov 12 \pi^3  (d-5)} + ...
\ee
Thus finally 
\be \la{c8}
\G_{6\, \infty} =  - {1\over 2^{7} \pi^3 \, (d-5) }    \int d^6x\ \te  \big[ {9\over 5} (DF)^2-{19\over 15} F^3 \big]   \ . 
\ee 
Using that   ${1\over d-5 } \to -\log \Lambda$  and that $C_2=2$  in the $SU(2)$ case    this can be written   also  as 
\be \la{c9}
\G_{6\, \infty} =   {1\over 2^{8} \pi^3  }  C_2  \log \Lambda  
 \int d^6x\ \te  \big[ {9\over 5} (DF)^2-{19\over 15} F^3 \big]   \ , 
\ee 
which is   consistent with \rf{820},\rf{82},\rf{822},\rf{823}  for  the same   values of the  coefficients 
\be
\beta_2 = - 27 \ , \qquad \ \  \beta_3 = -57 \ ,  \ee
as found in   section 4. 

\renewcommand{\theequation}{D.\arabic{equation}}
\setcounter{equation}{0}
\section{Non-abelian chiral 4d vector model}

Here we consider  a  4d analog of the  6d  self-dual $B$-field model \rf{21}  where  the  tensor   field $B_{ij}$ 
is replaced by a vector $B_i$  (with $i=1,2,3$)    which 
is coupled to a  gauge  field   living in  a 3d    subspace (or a ``defect"): 
\be\la{d1} 
S_4 = \int dx_4 d^3 x ~  \big[       (\partial_4   B^a_i )^2   + i m\,   \epsilon_{ijk}  B^a_i  D_k    B^a_j  \big]\ , 
 \qquad \qquad   D_k B^a_i=  \del_k B^a_i +    f^{abc} A^b_k B^c_i \ .
\ee
We assume that  $B_i$ is in adjoint representation   and depends on all 4 coordinates while 
$A^a_i$  depends only on 3  coordinates $x_i$.  
Here  we  can not  have $\del_4$   in the  $\eps_{ijk}$   term  as otherwise this term vanishes, so  we 
need to introduce a    mass parameter $m$ to balance the dimensions.  
The action is invariant under the  local symmetry  (with $U=U(x_i)$, cf. \rf{13}) 
\be
B'_i = UB_iU^{-1}\ , \qquad 
\qquad
A'_i = U  (A_i + \del_i)  U^{-1} \ .
\ee
Integrating out the $B_i$-field  one should  then get a gauge-invariant effective action $\G(A)$  depending on 
the 3d field $A_i$. 
Since  the  classical  action is not parity-invariant,  $\G$ may contain  a  non-local  P-odd part. 

The analogs of the propagator \rf{46} and the  vertex \rf{477}   linear in $A$   here are ($p^2=p_i p_i$) 
\be\la{d2}
P_{ij}^{ab} =\delta^{ab}   P_{ij}(p_s, p_4) 
 =   \frac{\delta^{ab} }{m^2 p^2 + p_4^4}  \Big(   p_4^2 \delta_{ij} - m \epsilon_{ijk} p_k  +\frac{m^2}{p_4^2} p_i p_j \Big) 
 , \ \ \ \ \ \  V_{ijk}^{abc} = - m \epsilon_{ijk} f^{abc}\ 
\ee
To compute  the effective action we  use  dimensional regularization in a 3d variant of the 4d helicity scheme, in which all the numerator algebra is carried out in   3 dimensions  and then the remaining scalar 
momentum integrals   are done in $d=3 - 2\eps$ dimensions (with $p_4$ integral treated as 1-dimensional one). 

The one-loop two-point function of the external  $A^a_i$-field  appearing in the $A^2$ term 
of the effective action  may be written as ($k_s$ is  an external 3-momentum)
\be
i\Pi_{ij}^{ab} (k_s)  &=&  \int \frac{d p_4}{2\pi} \frac{d^d p}{(2\pi)^d} \ V_{mn i}^{cda} \,  P_{nr}^{de}(p_s, p_4) \,  V_{rq j}^{efb} P_{qm}^{fc}  (p_s+k_s, p_4) = i \delta^{ab} \Pi_{ij} (k_s) \ ,
\cr
i \Pi_{ij} (k_s)  &=&-m^2 C_2   \epsilon_{imn  } \epsilon_{jpr }  \int \frac{d p_4}{2\pi} \frac{d^d p}{(2\pi)^d} \ P_{nr}(p_i, p_4) P_{qm}(p_i+k_i, p_4)\ . \la{d4}
\ee
Using the identity \rf{000} with $n=1$ (which is equivalent to  $\int \frac{d^d q}{(2\pi)^d} \frac{\partial}{\partial q^\mu} \frac{q_\mu}{q^2+X}=0$)  and introducing the Feynman  parameter $y$ 
leads to
\be
&&i\Pi_{ij}(k_s) =-m^2 C_2  \int_0^1 dy
 \int \frac{d p_4}{2\pi} \frac{d^d p}{(2\pi)^d} \ \frac{N_{ij}(k_s, p_4, y)}{p_4^4 \big[m^2 p^2 + p_4^4 + m^2 y(1-y) k^2\big]^2} \ , 
\\
&&N_{ij} = m^2 y(1-y) (4 p_4^4 + m^2 k^2) (k^2 \delta_{ij}  - k_ik_j) -
 2m \tfrac{d-1}{d-2}\,   p_4^2 \big[p_4^4 + m^2 y(1-y) k^2\big]
\epsilon_{ijs} k_s\ . 
\nonumber
\ee
Performing the $p_i$ integral using \rf{104}  gives 
\be\la{d5}
i\Pi_{ij} =-m^{2-d} C_2  \frac{\Gamma(2-d/2)}{(4\pi)^{d/2} }  \int_0^1 dy \int \frac{d p_4}{2\pi} 
\frac{N_{ij} (k_s, p_4, y)}{p_4^4\big[ p_4^4 + m^2 y(1-y) k^2\big]^{2-d/2}} \ .
\ee
To  integrate over  $p_4$  we first  change the  variable to $\mu$   as    $p_4 = \mu [m^2 y(1-y) k^2]^{1/4}$
and then use \rf{001}. Integrating  over $y$   we finally  obtain
\be\la{d6}
i\Pi_{ij} = -\tfrac{1}{30\pi}  \big({m^2\ov k^2}\big)^{1/4}(k^2 \delta_{ij}  - k_ik_j)  - \tfrac{1}{6\pi}  \big(m^2 k^2\big)^{1/4}\epsilon_{ijr} k_r \ . 
\ee
In contrast to the 6d case in \rf{48}--\rf{412}, here  the 
  $A^2$   term in the effective action is   UV  finite  and  contains two  3d gauge-invariant  non-local  structures:
P-even one   $ \int  F_{ij}  ({m^2\ov \del^2})^{1/4}  F_{ij} $  and   P-odd    one  
$\int  \,    \eps_{kij} A_k  ({m^2 \del^2})^{1/4}  F_{ij} $.



\begin{thebibliography}{99}

  
\bibitem{Saemann:2017zpd} 
  C.~Saemann and L.~Schmidt,
  ``Towards an M5-Brane Model I: A 6d Superconformal Field Theory,''
  J.\ Math.\ Phys.\  {\bf 59}, 043502 (2018)
  [arXiv:1712.06623].



\bibitem{Samtleben:2011fj} 
  H.~Samtleben, E.~Sezgin and R.~Wimmer,
  ``(1,0) superconformal models in six dimensions,''
  JHEP {\bf 1112}, 062 (2011)
  [arXiv:1108.4060]; 
  ``Six-dimensional superconformal couplings of non-abelian tensor and hypermultiplets,''
  JHEP {\bf 1303}, 068 (2013)
  [arXiv:1212.5199]. \ 
H.~Samtleben, E.~Sezgin, R.~Wimmer and L.~Wulff,
  ``New superconformal models in six dimensions: Gauge group and representation structure,''
  PoS Corfu {\bf 2011}, 071 (2011)
  [arXiv:1204.0542].
 
\bibitem{Lambert:2010wm} 
  N.~Lambert and C.~Papageorgakis,
  ``Nonabelian (2,0) Tensor Multiplets and 3-algebras,''
  JHEP {\bf 1008}, 083 (2010)
  [arXiv:1007.2982 [hep-th]].

  
\bibitem{Chu:2012um} 
  C.-S.~Chu and S.-L.~Ko,
  ``Non-abelian Action for Multiple Five-Branes with Self-Dual Tensors,''
  JHEP {\bf 1205}, 028 (2012)
  [arXiv:1203.4224 [hep-th]].

\bibitem{Bonetti:2012st} 
  F.~Bonetti, T.~W.~Grimm and S.~Hohenegger,
  ``Non-Abelian Tensor Towers and (2,0) Superconformal Theories,''
  JHEP {\bf 1305}, 129 (2013)
  [arXiv:1209.3017 [hep-th]].




\bibitem{Ho:2011ni} 
  P.-M.~Ho, K.-W.~Huang and Y.~Matsuo,
  ``A Non-Abelian Self-Dual Gauge Theory in 5+1 Dimensions,''
  JHEP {\bf 1107}, 021 (2011)
  [arXiv:1104.4040].




\bibitem{Ho:2012nt} 
  P.-M.~Ho and Y.~Matsuo,
  ``Note on non-Abelian two-form gauge fields,''
  JHEP {\bf 1209}, 075 (2012)
  [arXiv:1206.5643].
``Aspects of Effective Theory for Multiple M5-Branes Compactified On Circle,''
  JHEP {\bf 1412}, 154 (2014)
  [arXiv:1409.4060].

\bibitem{Huang:2012tu} 
  K.-W.~Huang,
  ``Non-Abelian Chiral 2-Form and M5-Branes,''
  arXiv:1206.3983.

\bibitem{Fradkin:1982kf} 
  E.~S.~Fradkin and A.~A.~Tseytlin,
  ``Quantum Properties of Higher Dimensional and Dimensionally Reduced Supersymmetric Theories,''
  Nucl.\ Phys.\ B {\bf 227}, 252 (1983).

\bibitem{Frampton:1983ah} 
  P.~H.~Frampton and T.~W.~Kephart,
  ``Explicit Evaluation of Anomalies in Higher Dimensions,''
  Phys.\ Rev.\ Lett.\  {\bf 50}, 1343 (1983)
  Erratum: [Phys.\ Rev.\ Lett.\  {\bf 51}, 232 (1983)].

\bibitem{AlvarezGaume:1983ig} 
L.~Alvarez-Gaume and E.~Witten, 
``Gravitational Anomalies,"
Nucl.  Phys.  B {\bf 234} (1984) 269

\bibitem{Bastianelli:1989hi} 
  F.~Bastianelli and P.~van Nieuwenhuizen,
  ``Gravitational Anomalies From the Action for Selfdual Antisymmetric Tensor Fields in (4k+2)-dimensions,''
  Phys.\ Rev.\ Lett.\  {\bf 63}, 728 (1989).


\bibitem{Beccaria:2017aqc} 
  M.~Beccaria and A.~A.~Tseytlin,
  ``Partition function of free conformal fields in 3-plet representation,''
  JHEP {\bf 1705}, 053 (2017)
  [arXiv:1703.04460].


\bibitem{Henneaux:1988gg} 
  M.~Henneaux and C.~Teitelboim,
  ``Dynamics of Chiral (Selfdual) $P$ Forms,''
  Phys.\ Lett.\ B {\bf 206}, 650 (1988).
  
\bibitem{Schwarz:1993vs} 
  J.~H.~Schwarz and A.~Sen,
  ``Duality symmetric actions,''
  Nucl.\ Phys.\ B {\bf 411}, 35 (1994)
  [hep-th/9304154].



\bibitem{Perry:1996mk} 
  M.~Perry and J.~H.~Schwarz,
  ``Interacting chiral gauge fields in six-dimensions and Born-Infeld theory,''
  Nucl.\ Phys.\ B {\bf 489}, 47 (1997)
  [hep-th/9611065].


\bibitem{Pasti:1996vs} 
  P.~Pasti, D.~P.~Sorokin and M.~Tonin,
  ``On Lorentz invariant actions for chiral p forms,''
  Phys.\ Rev.\ D {\bf 55}, 6292 (1997)
  [hep-th/9611100].


\bibitem{Belov:2006jd} 
  D.~Belov and G.~W.~Moore,
  ``Holographic Action for the Self-Dual Field,''
  [hep-th/0605038].



\bibitem{Gilkey:1975iq} 
  P.~B.~Gilkey,
  ``The Spectral geometry of a Riemannian manifold,''
  J.\ Diff.\ Geom.\  {\bf 10}, no. 4, 601 (1975).

\bibitem{Bastianelli:2000hi} 
  F.~Bastianelli, S.~Frolov and A.~A.~Tseytlin,
  ``Conformal anomaly of (2,0) tensor multiplet in six-dimensions and AdS/CFT correspondence,''
  JHEP {\bf 0002}, 013 (2000)
  [hep-th/0001041].

\bibitem{Osborn:2015rna} 
  H.~Osborn and A.~Stergiou,
  ``Structures on the Conformal Manifold in Six Dimensional Theories,''
  JHEP {\bf 1504}, 157 (2015)
  [arXiv:1501.01308].
  
  
  
\bibitem{Ivanov:2005qf} 
  E.~A.~Ivanov, A.~V.~Smilga and B.~M.~Zupnik,
  ``Renormalizable supersymmetric gauge theory in six dimensions,''
  Nucl.\ Phys.\ B {\bf 726}, 131 (2005)
  [hep-th/0505082].
 A.~V.~Smilga,
  ``Chiral anomalies in higher-derivative supersymmetric 6D theories,''
  Phys.\ Lett.\ B {\bf 647}, 298 (2007)
  [hep-th/0606139].

\bibitem{Buchbinder:2017ozh} 
 I.~L.~Buchbinder, E.~A.~Ivanov, B.~S.~Merzlikin and K.~V.~Stepanyantz,
  ``One-loop divergences in 6D, $ \mathcal{N} $ = (1, 0) SYM theory,''
  JHEP {\bf 1701}, 128 (2017) 
  [arXiv:1612.03190];
  ``Supergraph analysis of the one-loop divergences in $6D$, ${\cal N} = (1,0)$ and ${\cal N} = (1,1)$ gauge theories,''
  Nucl.\ Phys.\ B {\bf 921}, 127 (2017)
  [arXiv:1704.02530].
  E.~Ivanov,
  ``Classical and quantum superfield invariants in $\mathcal N = (1, 1), 6D$ SYM theory,''
  J.\ Phys.\ Conf.\ Ser.\  {\bf 965}, no. 1, 012021 (2018).




\bibitem{Bhattacharya:2016ydi} 
  J.~Bhattacharya and A.~E.~Lipstein,
  ``6d Dual Conformal Symmetry and Minimal Volumes in AdS,''
  JHEP {\bf 1612}, 105 (2016)
  [arXiv:1611.02179 [hep-th]].


\bibitem{Manvelyan:2000ea} 
  R.~Manvelyan and A.~C.~Petkou,
  ``The Trace anomaly of the (2,0) tensor multiplet in background gauge fields,''
  JHEP {\bf 0006}, 003 (2000)
  [hep-th/0005256].
  
  

  
\bibitem{Schwarz:1979ae} 
  A.~S.~Schwarz,
  ``The Partition Function of a Degenerate Functional,''
  Commun.\ Math.\ Phys.\  {\bf 67}, 1 (1979).
A.~S.~Schwarz and Y.~S.~Tyupkin,
  ``Quantization Of Antisymmetric Tensors And Ray-singer Torsion,''
  Nucl.\ Phys.\ B {\bf 242}, 436 (1984).


\bibitem{Beccaria:2014jxa} 
  M.~Beccaria, X.~Bekaert and A.~A.~Tseytlin,
  ``Partition function of free conformal higher spin theory,''
  JHEP {\bf 1408}, 113 (2014)
  [arXiv:1406.3542].

\end{thebibliography}
\end{document}